\documentclass[12pt,a4paper]{article}
\usepackage{graphicx}
\usepackage{setspace}
\usepackage{array}
\let\oldtabular=\tabular
\def\tabular{\tiny \oldtabular}
\newcolumntype{M}[1]{>{\raggedright}m{#1}}
\begin{document}
\title{mARC: Memory by Association and Reinforcement of Contexts}

\author{
Norbert Rimoux and Patrice Descourt
\\
Marvinbot S.A.S, France
}

\maketitle

\begin{abstract}

This paper introduces the memory by Association and Reinforcement of 
Contexts (mARC). mARC is a novel data modeling technology rooted in the 
second quantification formulation of quantum mechanics. It is an 
all-purpose incremental and unsupervised data storage and retrieval 
system which can be applied to all types of signal or data, structured 
or unstructured, textual or not. mARC can be applied to a wide range of 
information classification and retrieval problems like e-Discovery or 
contextual navigation. It can also formulated in the artificial life 
framework a.k.a Conway "\,Game Of Life" Theory. In contrast to Conway 
approach, the objects evolve in a massively multidimensional space. In 
order to start evaluating the potential of mARC we have built a 
mARC-based Internet search engine demonstrator with contextual 
functionality. We compare the behavior of the mARC demonstrator with 
Google search both in terms of performance and relevance. In the study 
we find that the mARC search engine demonstrator outperforms Google 
search by an order of magnitude in response time while providing more 
relevant results for some classes of queries.
\end{abstract}

\section{Introduction}

At the onset of 20$^{th}$ century, it was generally believed that a 
complex system was the sum of its constituents. Furthermore, each 
constituent could be analyzed independently of the others and 
reassembled together to bring the whole system back.

Since the advent of quantum physics with Dirac, Heinsenberg, 
Schrödinger, Wigner, etc. and the debate about the incompleteness of the 
probabilistic formulation of quantum mechanics which arose between 
Einstein and the Copenhagen interpretation of quantum physics led by 
Niels Bohr in 1935, the paradigm enounced in the beginning of this 
paragraph has been seriously questioned.

The EPR thought experiment $[$Einstein1935$]$ at the heart of this 
debate opened the path for Bell inequalities which concern measurements 
made by observers on pairs of particles that have interacted and then 
separated. In quantum theory, such particles are still strongly 
entangled irrespective of the distance between them. According to 
Einstein's local reality principle (due to the finiteness of the speed 
of light), there is a limit to the correlation of subsequent 
measurements of the particles.

This experiment opened the path to Aspect's experiments between 1980 and 
1982 which showed a violation of Bell inequalities and proved the 
non-local and non-separable orthodox formulation of quantum theory 
without hidden variables $[$Aspect1981, Aspect1982$]$.

After this holistic shift in the former Newtonian and Cartesian 
paradigms, Roger Penrose and others have argued that quantum mechanics 
may play an essential role in cognitive processes $[$Penrose1989, 
Penrose1997$]$. This contrasts with most current mainstream biophysics 
research on cognitive processes where the brain is modeled as a neural 
network obeying classical physics. We may so wonder if, for any 
artificial intelligence system seen as a complex adaptive system (CAS), 
quantum entanglement should not be an inner feature such as emergence, 
self-organization and co-evolution. 

$[$Rijsbergen2004$]$ further demonstrates that several models of 
information retrieval (IR) can be expressed in the same framework used 
to formulate the general principles of quantum mechanics.

Building on this principle, we have designed and implemented a complex 
adaptive system, the memory by Association and Reinforcement of Contexts 
(mARC) that can efficiently tackle the most complex information 
retrieval tasks.

The remainder of this paper is organized as follows: Section 2 describes 
the current approaches to machine learning and describes related work. 
Section 3 describes mARC. Section 4 compares the performance and 
relevance of results of a mARC-based search demonstrator and Google 
search. Finally, section 5 draws some conclusions and describes future 
work.

\section{Current Approaches}

\subsection{Text Mining}

Text mining covers a broad range of related topics and algorithms for 
text analysis. It spans many different communities among which: natural 
language processing, named entity recognition, information retrieval, 
text summarization, dimensionality reduction, information extraction, 
data mining, machine learning (supervised, unsupervised and 
semi-supervised) and many applications domains such as the World Wide 
Web, biomedical science, finance and media industries.

The most important characteristic of textual data is that it is sparse 
and high dimensional. A corpus can be drawn from a lexicon of about one 
hundred thousand words, but a given text document from this corpus may 
contain only a few hundred words. This characteristic is even more 
prominent when the documents are very short (tweets, emails, messages on 
a Facebook wall, etc.).

While the lexicon of a given corpus of documents may be large, the words 
are typically correlated with one another. This means that the number of 
concepts (or principal components) in the data is much smaller than the 
feature space. This advocates for the careful design of algorithms which 
can account for word correlations.

Mathematically speaking, a corpus of text documents can be represented 
as a huge, massively high-dimensional, sparse term/document matrix. Each 
entry in this matrix is the normalized frequency of a given term in the 
lexicon in a given document. \textit{Term frequency-inverse document 
frequency} (TF-IDF) is currently the most accurate and fastest 
normalization statistic that can take into account the proper 
normalization between the local and global importance of a given word 
inside a document with respect to the corpus. Note, however, that it has 
been shown recently that binary weights give more stable indicators of 
sentence importance than word probability and TF-IDF in topic 
representation for text summarization $[$Gupta2007$]$.

Because of the huge size and the sparsity of the text/document matrix, 
all correlation techniques suffer from the curse of dimensionality. 
Moreover, the variability in word frequencies and document lengths also 
creates a number of issues with respect to document representation and 
normalization. These are critical to the relevance, efficiency and 
scalability of state of the art classification, information extraction, 
or statistical machine learning algorithms.

Textual data can be analyzed at different representation levels. The 
primary and most widely investigated representation in practical 
applications is the \textit{bag of words} model. However, for most 
applications, being able to represent text information \textit{
semantically} enables a more meaningful analysis and text mining. This 
requires a major shift in the canonical representation of textual 
information to a representation in terms of named entities such as 
people, organizations, locations and their respective relations 
$[$Etzioni2011$]$. Only the proper representation of explicit and 
implicit contextual relationships (instead of a bag of words) can enable 
the discovery of more interesting patterns. $[$Etzioni2011$]$ 
underscores the urgent need to go beyond the keyword approximation 
paradigm. Looking at the fast expanding body scientific literature from 
which people struggle to make sense, gaining insight into the semantics 
of the encapsulated information is urgently needed $[$Lok2010$]$.

Unfortunately, state of the art methods in natural language processing 
are still not robust enough to work well in unrestricted heterogeneous 
text domains and generate accurate semantic representations of text. 
Thus, most text mining approaches currently rely on the word-based 
representations, especially the bag of words model. This model, despite 
losing the positioning and relational information in the words, is 
generally much simpler to deal with from an algorithmic point of view 
$[$Aggarwal2012$]$. 

Although statistical learning and language have so far been assumed to 
be intertwined, this theoretical presupposition has rarely been tested 
empirically $[$Misyak2012$]$. As emphasized by Clark in $[$Clark1973$]$, 
\textit{current investigators of words, sentences, and others language 
materials almost never provide statistical evidence that their findings 
generalize beyond the specific sample of language materials they have 
chosen}. Perhaps the most frustrating aspect of statistical language 
modeling is the contrast between our intuition as speakers of natural 
languages and the over-simplistic nature of our most successful models 
$[$Rosenfeld2000$]$.

Supervised learning methods exploit training data which is manually 
created, annotated, tagged and classified by human beings in order to 
train a classifier or regression function that can be used to compute 
predictions on new data. This learning paradigm is largely in use in 
commercial machine language processing tools to extract information and 
relations about facts, people and organizations. This requires large 
training data sets and numerous human annotators and linguists for each 
language that needs to be processed.

The current methods comprise rules-based classifiers, decision trees, 
nearest neighbors classifiers, neural networks classifiers, maximal 
margins classifiers (like support vector machines) and probabilistic 
classifiers like \textit{conditional random fields} (CRF) for name 
entity recognition, \textit{Bayesian networks} (BN) and Markov 
processes such as \textit{Hidden Markov Models} (HMMs) (currently used 
in part-of-speech tagging and speech recognition), \textit{
maximum-entropy Markov models} (MEMMs), and \textit{Markov Random 
Fields}. CRF has been applied to a wide variety of problems in natural 
language processing, including POS tagging $[$Lafferty2001$]$, shallow 
parsing $[$Sha2003$]$, and named entity recognition $[$McCallum2003$]$ 
as an alternative to the related HMMs.

Many statistical learning algorithms treat the learning task as a 
\textit{sequence labeling} problem. Sequence labeling is a general 
machine learning technique. It has been used to model many natural 
language processing tasks including part-of-speech tagging, chunking and 
named entity recognition. It assumes we are given a sequence of 
observations. Usually each observation is represented as feature vectors 
which interact through feature functions to compute conditional 
probabilities.

As a simple example, let us consider $x_{1:N}$ be a set of 
observations (e.g. words in a document), and $z_{1:N}$ the hidden 
labels (e.g. tags). Let us also assume that each observation can be 
expressed in terms of F features. A linear chain conditional random 
field de?nes the conditional probability that a given tag is associated 
with a document knowing that a given word has been observed as:

$p(z_{1:N}x_{1:N})=\frac{1}{Z}e^{\sum_{n=1}^{N}{\sum_{i=1}^{F}{\lambda 
_{i}}f_{i}(z_{n-1},z_{n},x_{1}:N,n)}} $\\

Z is just there to ensure that all the probabilities sum to one, i.e. it 
is a normalization factor. For example, we can de?ne a simple feature 
function which produces binary values: it is 1 if the current word is 
"\,John", and if the current state $z_{n}$ is "\,PERSON":

$$
f_{1}(z_{n-1},z_{n},x_{1:N},n)=\left\{
 \begin{array}{lr}
1 & if z_{n} ="PERSON" and x_{n} ="John" \\
0 & otherwise \\
\end{array}
\right.
$$

How this feature is used depends on its corresponding weight ?$_{1}$
. If ?$_{1}$ $>$ 0, whenever f$_{1}$ is active (i.e. we see the 
word John in the sentence and we assign it the tag PERSON), it increases 
the probability of the tag sequence z$_{1:N}$. This is another way 
of saying "\,the CRF model should prefer the tag PERSON for the word 
John". 

 A common way to assign a label to each observation is to model the joint 
probability as a Markov process\textbf{ }where the generation of a 
label or an observation is dependent only on one or a few previous 
labels and/or observations. This technique is currently extensively used 
in the industry. Although Markov chains are ef?cient at encoding local 
word interactions, the n-gram model clearly ignores the rich syntactic 
and semantic structures that constrain natural languages $[$Ming2012$]$. 
Attempting to increase the order of an n-gram to capture longer range 
dependencies in natural language immediately runs into the 
dimensionality curse $[$Bengio2003$]$.

Unfortunately, from a computational point of view, even if we restrict 
the process to be linear (depending only on one predecessor) the task is 
highly demanding in computational resources. The major di?erence between 
CRFs and MEMMs is that in CRFs the label of the current observation can 
depend not only on previous labels but also on future labels.

In mathematical graph theory terms, CRFs are undirected graph models 
while both HMMs and MEMMs are directed graph models. Usually, 
linear-chain CRFs are used for sequence labeling problems in natural 
language processing where the current label depends on the previous 
label and the next label as well as the observations. In linear-chain 
CRFs long-range features cannot be de?ned. General CRFs allow long-range 
features but are too expensive to perform exact inference. Sarawagi and 
Cohen have proposed \textit{semi-Markov conditional random ?elds} as a 
compromise $[$Saragawi2005$]$. In semi-Markov CRFs, labels are assigned 
to segments of the observation sequence and features can measure 
properties of these segments. Exact learning and inference on 
semi-Markov CRFs is thus computationally feasible and consequently 
achieves better performance than standard CRFs because they take into 
account long-range features.

HMM models have been applied to a wide variety of problems in 
information extraction and natural language processing, especially POS 
tagging $[$Kupiec1992$]$ and named entity recognition $[$Bikel1999$]$. 
Taking POS tagging as an example, each word is labeled with a tag 
indicating its appropriate part of speech, resulting in annotated text, 
such as: "\,$[$VB heat$]$ $[$NN water$]$ $[$IN in$]$ $[$DT a$]$ $[$JJ 
large$]$ $[$NN vessel$]$". Given a sequence of words, e.g. "\,heat water 
in a large vessel", the task is to assign a sequence of labels e.g. 
"\,VB NN IN DT JJ NN", for the words. HMM models determine the sequence 
of labels by maximizing a joint probability distribution computed from 
the manually annotated training data. In practice, Markov processes like 
HMM require independence assumptions among the random variables in order 
to ensure tractable inference.

The primary advantage of CRFs over HMMs is their conditional nature 
resulting in the relaxation of the independence assumption. However, the 
problem of exact inference in CRFs is nevertheless intractable. 
Similarly to HMMs, the parameters are typically learned by maximizing 
the likelihood of training data and need rely on iterative techniques 
such as iterative scaling $[$Lafferty2001$]$ and gradient-descent 
methods $[$Sha2003$]$.

All these models depend on multiple parameters to define the underlying 
prior probabilistic distributions used to generate the posterior 
distributions which describe the observed labeled data in order to infer 
classification on unlabeled data. Canonical well know and well-studied 
probability distributions like Gaussian, multinomial, Poisson, or 
Dirichlet are primarily used in these models. The paradigmatic 
mathematical formulation of these models in terms of "cost", "score" or 
"energy" functions rely on the maximization of the latter.

Unfortunately, these models are embedded in huge multi-dimensional 
spaces. Finding the set of parameters which actually minimize these 
functions is a combinatorial optimization problem and is known to be 
NP-hard. Heuristic algorithms to compute the parameters are fairly 
complex and difficult to implement $[$Teyssier2012$]$. Moreover, 
parameter estimation for the prior distribution functions is essentially 
based on conditional counting with various normalization and 
regularization smoothing schemes to correct for sparseness of a given 
occurrence in the observed and training data. These parameterization 
schemes greatly vary in the literature and there is no canonical or 
natural heuristic to determine them for each application domain.

The learning algorithms for these probabilistic models try to ?nd 
maximum-likelihood estimation (MLE) and maximum \textit{a posteriori} 
probability (MAP) estimators for the parameters in these models. Most of 
the time, no closed form solutions can be provided.

In order to be able to make predictions from these models, canonical 
learning schemes such as \textit{Expectation-Maximization} (EM) 
$[$Blei2003$]$ $[$Borman2004$]$, Gibbs sampling and \textit{Markov 
Chain Monte Carlo} are used extensively $[$Andrieu2003$]$. In recent 
years, the main research trend in this field has been in the context of 
two classes of text data:

\begin{itemize}
\item Dynamic Applications
\end{itemize}
The large amount of text data being generated by dynamic applications 
such as social networks or online chat applications has created a 
tremendous need for clustering streaming text. Such streaming 
applications must be applicable to text which is not very clean, as is 
often the case for social networks.

\begin{itemize}
\item Heterogeneous Applications
\end{itemize}
Text applications increasingly arise in heterogeneous applications in 
which the text is available in the context of links, and other 
heterogeneous multimedia data. For example, in social networks such as 
Flickr, clustering often needs to be applied. Therefore, it is critical 
to effectively adapt text-based algorithms to heterogeneous multimedia 
scenarios.

Unsupervised learning techniques do not require any training data and 
therefore no manual effort. The two main applications are clustering and 
topic modeling. The basic idea behind topic modeling is to create a 
probabilistic generative model for the text documents in the corpus. The 
main approach is to represent a corpus as a function of hidden random 
variables, the parameters of which are estimated using a particular 
document collection.

There are two basic methods for topic modeling: \textit{Probabilistic 
Latent Semantic Indexing} (PLSI) $[$Hofmann1999$]$ and \textit{Latent 
Dirichlet Allocation} (LDA) $[$Blei2004$]$. Supervised information 
extraction comprises \textit{Hidden Markov models}, \textit{
Conditional Random Fields} or \textit{Support Vector Machines}. 
These techniques are currently heavily in use in the machine learning 
industry. All these techniques require the preprocessing of documents 
through manual annotation. For domain-speci?c information extraction 
systems, the annotated documents have to come from the target domain. 
For example, in order to evaluate gene and protein name extraction, 
biomedical documents such as PubMed abstracts are used. If the purpose 
is to evaluate general information extraction techniques, standard 
benchmark data sets can be used. Commonly used evaluation data sets for 
named entity recognition include MUC $[$Grishman1996$]$, CoNLL-2003 
$[$Tjong2003$]$ and ACE $[$ACE$]$. For relation extraction, ACE data 
sets are usually used. Currently, state-of-the-art named entity 
recognition methods can achieve around 90\% of F-1 scores (geometric 
mean of precision and recall) when trained and tested on the same domain 
$[$Tjong2003$]$.

For relation extraction, state-of-the-art performance is lower than that 
of named entity recognition. On the ACE 2004 benchmark dataset, for 
example, the best F-1 score is around 77\% for the seven major relation 
types $[$LongHua2008$]$.

It is generally observed that person entities are easier to extract, 
followed by locations and then organizations. It is important to note 
that when there is a domain change, named entity recognition performance 
can drop substantially. There have been several studies addressing the 
domain adaptation problem for named entity recognition $[$Jiang2006$]$.

Another new direction is open information extraction, where the system 
is expected to extract all useful entity relations from a large, diverse 
corpus such as the World Wide Web. The output of such systems includes 
not only the arguments involved in a relation but also a description of 
the relation extracted from the text.

In $[$Banko2008$]$, Banko and Etzioni have introduced an un-lexicalized 
CRF-based method for open information extraction. This method is based 
on the observation that although di?erent relation types have very 
di?erent semantic meanings, there exists a small set of syntactic 
patterns that covers the majority of the semantic relation mentions. The 
method categorizes binary relationships using a compact set of 
lexico-syntactic patterns. The heuristics are designed to capture 
dependencies typically obtained via syntactic parsing and semantic role 
labeling. For example, a heuristic used to identify positive examples is 
the extraction of noun phrases participating in a subject verb-object 
relationship e.g. "\,$<$Einstein$>$ received $<$the Nobel Prize$>$ in 
1921." An example of a heuristic that locates negative examples is the 
extraction of objects that cross the boundary of an adverbial clause, 
e.g. "\,He studied $<$Einstein's work$>$ when visiting $<$Germany$>$".

The set of features used by CRF is largely similar to those used by 
state-of-the-art relation extraction systems. They include 
part-of-speech tags (predicted using a separately trained 
maximum-entropy model), regular expressions (e.g. detecting 
capitalization, punctuation, etc.), context words, and conjunctions of 
features occurring in adjacent positions within six words to the left 
and six words to the right of the current word. The Open IE system 
extracts different relationships with a precision of 88.3\% and a recall 
of 45.2\%. However, the CRF-based IE system (O-CRF) has a number of 
limitations, most of which are shared with other systems that perform 
extraction from natural language text. First, O-CRF only extracts 
relations that are \textit{explicitly} mentioned in the text; implicit 
relationships that could inferred from the text would need to be 
inferred from O-CRF extractions. Second, O-CRF focuses on relationships 
that are primarily word-based, and not indicated solely from punctuation 
or document-level features. Finally, relations must occur between entity 
names within the same sentence.

With the fast growth of textual data on the Web, we expect that future 
work on information extraction will need to deal with even more diverse 
and noisy text. Weakly supervised and unsupervised methods will play a 
larger role in information extraction. The various user-generated 
content on the Web such as Wikipedia articles will also become important 
resources to provide some kind of supervision for $[$Aggarwal2012$]$.

In some applications, prior knowledge may be available about the kinds 
of clusters available in the underlying data. This prior knowledge may 
take on the form of labels attached with the document which indicate its 
underlying topic.

Such knowledge can be very useful in creating signi?cantly more coherent 
clusters, especially when the total number of clusters is large. The 
process of using such labels to guide the clustering process is referred 
to as semi-supervised clustering. This form of learning is a bridge 
between the clustering and classi?cation problem, because it uses the 
underlying class structure, but is not completely tied down by the 
speci?c structure. As a result, this approach is applicable to both the 
clustering and classi?cation scenarios. The most natural way of 
incorporating supervision into the clustering process is partitional 
clustering methods such as \textit{k-means}. This is because 
supervision can be easily incorporated by changing the seeds in the 
clustering process $[$Aggarwal2004, Basu2002$]$. A number of 
probabilistic frameworks have also been designed for semi-supervised 
clustering $[$Nigam1998, Basu2004$]$. 

However real world applications in these fields currently lack scalable 
and robust methods for natural language understanding and modeling 
$[$Aggarwal2012$]$. For example, current information extraction 
algorithms mostly rely on costly, non-incremental, and time consuming 
supervised learning and generally only work well when sufficient 
structured and homogeneous training data is available. This requirement 
drastically restricts the practical application domains of these 
techniques $[$Aggarwal2012$]$.

All the models described above are computationally intensive. The 
e?ciency of the learning algorithms is always an issue, especially for 
large scale data sets which are quite common for text data. In order to 
deal with such large datasets, algorithms with linear or even sub-linear 
time complexity are required, for which parallelism can be used to speed 
up computation.

MapReduce $[$Dean2004$]$ is a \underline{programming model} for 
processing large data sets, and the name of an implementation of the 
model by \underline{Google}. MapReduce is typically used to do 
\underline{distribute computations} on \underline{clusters} of 
computers. Apache Hadoop (http://hadoop.apache.org) is an \underline{
open-source} implementation of MapReduce. It supports data-intensive 
\underline{distributed applications} and running these applications on 
large clusters of commodity hardware. The major algorithmic challenges 
in map-reduce computations involve balancing a multitude of factors such 
as the number of machines available for mappers/reducers, their memory 
requirements, and communication cost (total amount of data sent from 
mappers to reducers) $[$Foto2012$]$.

Figure 1 (taken from $[$Hockenmaier$]$) presents the training time for 
syntactic translation models using Hadoop. On the right, the benefit of 
distributed computation quickly outweighs the overhead of a MapReduce 
implementation on a 3-node cluster. However, on the left, we see that 
exporting the data to the distributed file system incurs cost nearly 
equal to that of the computation itself.

 Existing tools do not lend themselves to sophisticated data analysis at 
the scale many users would like $[$Maden2012$]$. Tools such as SAS, R, 
and Matlab support relatively sophisticated analysis, but are not 
designed to scale to datasets that exceed even the memory of a single 
computer. Tools that are designed to scale, such as relational DBMSs and 
Hadoop, do not support these algorithms out of the box. Additionally, 
neither DBMSs nor MapReduce are particularly efficient at handling high 
incoming data rates and provide little out-of-the-box support for 
techniques such as approximation, single-pass/sub linear algorithms, or 
sampling that might help ingest massive volumes of data.

\begin{figure}[h]
\centering
\includegraphics[width=14.00cm,height=6.00cm]{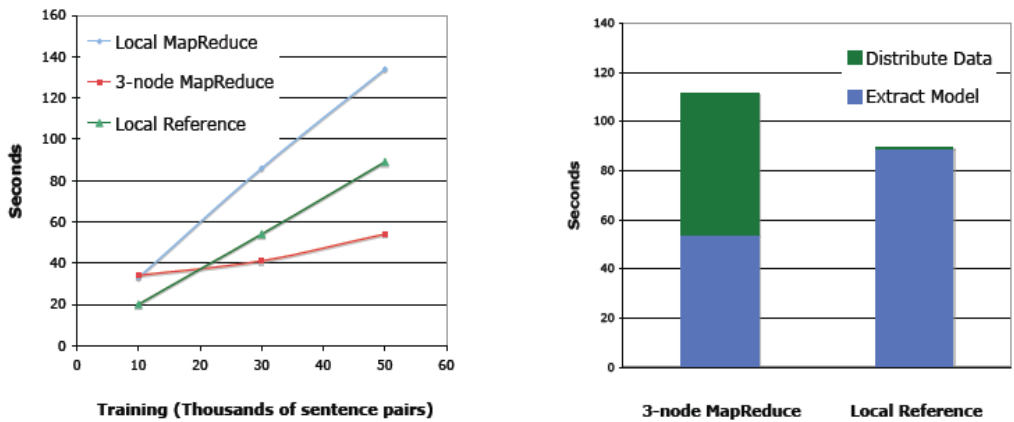}
\caption{MapReduce/Hadoop comparison on training data}
\end{figure}

Several research projects are trying to bridge the gap between 
large-scale data processing platforms such as DBMSs and MapReduce, and 
analysis packages such as SAS, R, and Matlab. These typically take one 
of three approaches: extend the relational model, extend the 
MapReduce/Hadoop model, or build something entirely different. In the 
relational camp are traditional vendors such as Oracle, with products 
like its Data Mining extensions, as well as upstarts such as Greenplum 
with its Mad Skills project. However, machine learning algorithms often 
require considerably sophisticated users, especially with regard to 
selecting features for training and choosing model structure (for 
instance, for regression or in statistical graphical models).

In the past two decades $[$DU2012$]$, most work in speech and language 
processing has used "\,shallow" models which lack multiple layers of 
adaptive nonlinear features. Current speech recognition systems, for 
example, typically use \textit{Gaussian mixture models} (GMMs), to 
estimate the observation (or emission) probabilities of \textit{hidden 
Markov models} (HMMs) $[$Singh2012$]$. GMMs are generative models that 
have only one layer of latent variables. Instead of developing more 
powerful models, most of the research has focused on finding better ways 
of estimating the GMM parameters so that error rates are decreased or 
the margin between different classes is increased. The same observation 
holds for natural language processing (NLP) in which maximum entropy 
(MaxEnt) models and conditional random fields (CRFs) have been popular 
for the last decade. Both of these approaches use shallow models whose 
success largely depends on the use of carefully handcrafted features.

Shallow models have been effective in solving many simple or 
well-constrained problems, but their limited modeling power can cause 
difficulties when dealing with more complex real-world applications. For 
example, a state-of-the-art GMM-HMM based speech recognition system that 
achieves less than 5\% word error rate (WER) on read English may exceed 
15\% WER on spontaneous speech collected under real usage scenarios due 
to variations in environment, accent, speed, co-articulation, and 
channel.

Existing deep models like hierarchical HMMs or \textit{Higher Order 
Conditional Random Fields} (HRCRFs) and multi-level detection-based 
systems are quite limited in exploiting the full potential that deep 
learning techniques can bring to advance the state of the art in speech 
and language processing.

\subsection{Recommender Systems}

Recommender systems apply data mining techniques and prediction 
algorithms to the prediction of users' interest on information, products 
and services among vast amounts of available items (e.g. Amazon, 
Netflix, movieLens, and VERSIFY). The growth of information on the 
Internet as well as the number of website visitors add key challenges to 
recommender systems $[$Almazro2010$]$. Two recommendation techniques are 
currently extensively used in the industry $[$Zhou2012$]$: content based 
filtering (CBF) and collaborative filtering (CF). The content-based 
approach recommends items whose content is similar to content that the 
user has previously viewed or selected. The CBF systems relies on an 
extremely variable specific representation of items features, e.g. for a 
movie CBF, each film is featured by genre, actors, director, etc.

Knowledge-based recommendation attempts to suggest objects based on 
inferences about user needs and preferences. In some sense, all 
recommendation techniques could be described as doing some kind of 
inference. Knowledge-based approaches are particular in that they have 
functional knowledge: they have knowledge about how a particular item 
meets a particular user need and can therefore reason about the 
relationship between a need and a possible recommendation. The user 
profile can be any knowledge structure that supports this inference. In 
the simplest case, as in Google, it may simply be the query that the 
user has formulated. In others, it may be a more detailed representation 
of the user needs $[$Burke2002$]$.

The features retained to feed recommendation systems are generally 
created by human beings. Building the set of retained features is of 
course very time consuming, expensive and highly subjective. This 
subjectivity may impair the classification and recommendation efficiency 
of the system.

Collaborative filtering (CF) systems collect information about users by 
asking them to rate items and make recommendations based on the highly 
rated items by users with similar taste. CF approaches make 
recommendations based on the ratings of items by a set of users 
(neighbors) whose rating profiles are most similar to that of the target 
user. In contrast to CBF systems, CF systems rely on the availability of 
user profiles which capture past ratings and do not require any human 
intervention for tagging content because item knowledge is not required. 
CF is the most widely used approach for building recommender systems. It 
is currently used by Amazon to recommend books, CDs and many other 
products. Some systems combine CBF and CF techniques to improve and 
enlarge the capabilities of both approaches.

The quality and availability of user profiles is critical to the 
accuracy of recommender system. This information can be implicitly 
gathered by software agents that monitor user activities such as real 
time click streams and navigation patterns. Other agents collect 
explicit information about user interest from the ratings and items 
selected. Both the explicit and implicit methods have strengths and 
weaknesses. On the one hand, explicit interactions are more accurate 
because they come directly from the user but require a much greater user 
involvement. On the other hand, implicit monitoring requires little or 
no burden on the user but inferences drawn from the user interaction do 
not faithfully measure user interests. Hence, user profiles are often 
difficult to obtain and their quality is also both hard to ensure and 
assess.

Current existing user profiling for recommender systems is mainly using 
user rating data. Hundreds of thousands of items and users are 
simultaneously involved in a recommender system, while only a few items 
are viewed, rated or selected by users. Sarwar \textit{et al}. 
$[$Sarwar2001$]$ have reported that the density of the available ratings 
in commercial recommender systems is often less than 1\%. Moreover, new 
users start with a blank profile without selecting or rating any items 
at all. These situations are commonly referred to as data sparseness and 
cold start problem. The current recommender algorithms are impeded by 
the sparseness and cold start problems.

With the increased importance of recommender systems in e-commerce and 
social networks, the deliberate injection of false user rating data has 
also intensified. A simple, yet effective attack on recommender systems 
is to deliberately create a large number of fake users with pseudo 
ratings to favor or disfavor a particular product. With such fake 
information, user profile data can become unreliable.

In summary, without sufficient knowledge about users, even the most 
sophisticated recommendation strategies are not be able to make 
satisfactory recommendations. The cold start, data sparseness and 
malicious ratings are outstanding problems for user profiling. These 
make user profiles the weakest link in the whole recommendation process.

To tackle these issues, social recommender systems use user-generated 
(created) contents which comprise various forms of media and creative 
works as written, audio, visual and combined created by users explicitly 
and pro-actively $[$Pu2012$]$. Another path to improve performance, 
combine the above techniques in so-called hybrid recommenders 
$[$Burke2002$]$.

\section{mARC}

\subsection{Principle of Operation}

The Memory by Association and Reinforcement of Contexts (mARC) is an 
incremental, unsupervised and adaptive learning and pattern recognition 
system. Its ground principles allow the automatic detection and 
recognition of different types of patterns which are contextually 
linked.

mARC is built upon the premises introduced in $[$USP2004$]$. Companies 
such as IBM, Seagate Technology, and Nuance Communications have 
referenced this work in their patents and products.

Unlike systems such as \textit{feed-forward} or \textit{recurrent 
neural networks} and \textit{guided propagation networks} (GPN), mARC 
does not require a large memory space to run and has a fast response 
time. Furthermore, artificial neural network systems require the weights 
to be known before the network can be deployed and their capability to 
recognize patterns in known systems are limited $[$Papert1969$]$. 

The core of mARC is a fractal self-organized network whose basic element 
is called a \textit{cell}. A cell is an abstract structure used to 
encode any pattern from the incoming signal or any pattern from feedback 
signal inside the network.

The fractal structure naturally emerges as a consequence of the building 
and learning processes taking place inside the whole network.

A mARC server consists of the following elements:

\begin{itemize}
\item A networking socket.
\item A reading head or \textit{sensorial layer}.
\item A highly-optimized integrated binary database for fast storage and 
indexing of the input signal.
\item A core referred to as \textit{knowledge}.
\item An application programming interface (API) which allows 
interaction with the core.
\end{itemize}
 The network is initially empty, i.e. it does not contain any cells. At 
the top of the network is a reading head which reads a causal 
one-dimensional numeric input signal. 

In the input signal flow, the relative event time (causal appearance) 
describes the position of an event relative to another event. This can 
be seen as a relative time quantification between two event occurrences. 
As an example, if the incoming signal flow is 838578, sampled as 
83|85|78 coding for the word \textit{SUN} in extended ASCII, the event 
\textit{U} appears after event \textit{S} and prior to event 
\textit{N}. This is the relative time quantification of the event 
\textit{U} in the context of the pattern, the word \textit{SUN} in 
this example. In general, events are handled at the cell level and 
relative event times are handled at the global network level. 

The mARC implementation described in this paper is calibrated to sample 
the input signal byte-wise. In other words, it interprets the input 
signal as extended ASCII. As the ASCII input signal is presented to the 
network, it is transcoded into cells in the network. The network grows 
according to the input signal pattern. The input signal is composed of 
basic components or events in some order of occurrence linked by unknown 
causal patterns.

If a cell matching the basic component is found, that cell is reinforced 
(reinforcement learning and recognition) in the network. If a cell does 
not exist, a new cell is created to hold the basic component. As the 
cells are propagated in the network, a path encoding the pattern is 
automatically inserted in the structure of the network.

The learning and building processes are deeply intertwined. At any given 
time, the network contains a plurality of cell structures enabled to be 
linked to parent cells, cousin cells, and children cells in what we 
refer to as a "\textit{tricel" physical structure}. Each cell 
controls its own behavioral functions and transfers control to the next 
linked cells (self-signal forward and backward internal and external 
propagation).

A cell may have an attribute type of \textit{termination} or \textit{
glue}. A termination attribute marks the end of a learned and 
recognized segment in a pattern. A glue attribute indicates that a cell 
is an embedded event in a pattern. That is, a termination attribute 
typically marks an end of a significant recognized pattern. The 
termination cell may also include a link to another sub-network where 
related patterns are stored. These networks further aid in identifying 
an input pattern.

In other words, the network itself is the resultant of deeply 
inter-related and interacting layers of cells which draw a huge and 
massively multi-dimensional knowledge non-directed graph in the 
mathematical sense.

\subsection{The mARC Programming Model}

Interacting with mARC is performed via an application programming 
interface (API). The purpose of the API is to translate the internal 
structures of the mARC knowledge into object collections which are 
easier to handle procedurally.

For a text signal-oriented mARC, objects are typically words, compound 
expressions or phrases. The API automatically translates the inner 
contextual information of the mARC knowledge into weighted values for 
each object in a set according to their generality and activity with 
respect to the whole knowledge.

We distinguish two kinds of sets. We call \textit{genuine} or \textit{
canonical context}, a set of patterns which are genuinely correlated 
by the core. We call \textit{generic context} a context which is 
manually created using the API.

For example, let us assume that we want to probe the knowledge about the 
pattern \textit{bee}. The API contains a specific command for this. 
We instruct the API to build an empty context and put the pattern 
\textit{bee} in it. For now, the context has no genuine meaning with 
respect to the knowledge. The resulting context is generic. The API 
allows us to retrieve the genuine contexts from this generic context.

The genuine contexts are learned by the knowledge automatically from the 
corpus which has been submitted to it. The API allows the manipulation 
of the genuine contexts to perform true contextual analysis from the 
knowledge extracted from a corpus. Each element of a context (generic or 
genuine) is associated with two numerical values or \textit{weights }
internally computed from the knowledge: the \textit{generality} and 
the \textit{activity}. The activities of each element in a generic 
context have no meaning; they are arbitrarily fixed by the user. The 
activities are reevaluated by the knowledge once the genuine contexts 
issued from the generic context are retrieved from the knowledge.

The generality of an element inside a genuine context is a numerical 
estimate of the corresponding human notion with respect to the corpus 
which has been learned. The activity of an element inside a genuine 
context is an algebraic measure of the intensity of the coupling of each 
constituent of this context with respect to all the connections in the 
knowledge. The strength of this coupling is proportional to the number 
of connections between an element and its corresponding linked elements 
in the knowledge network.

\section{Key Differentiators}

mARC presents a number of key differentiators compared to other data 
processing and querying technologies:

\newcounter{numberedCntCC}
\begin{enumerate}
\item Independence from the data
\setcounter{numberedCntCC}{\theenumi}
\end{enumerate}
mARC is independent from the nature of the input signal. For example, 
mARC extracts contexts from textual data independently of the language 
the text is written in.\textbf{ }\textit{mARC handles any textual 
data as a numerical signal}. In essence, it is therefore a\textbf{ }
\textit{general numerical signal analysis processing unit}. Right 
now, it is restricted to handle byte-wise sampled signal i.e. Latin 9 or 
extended ASCII.

\begin{enumerate}
\setcounter{enumi}{\thenumberedCntCC}
\item Access time
\setcounter{numberedCntCC}{\theenumi}
\end{enumerate}
Access to contextual data is at least one order of magnitude faster than 
access to data using classical SQL-based language.

\begin{enumerate}
\setcounter{enumi}{\thenumberedCntCC}
\item Noise filtering and error correction
\setcounter{numberedCntCC}{\theenumi}
\end{enumerate}
Assuming enough contextual information is available, useful data can be 
filtered from noise. Data can also be reconstructed by mARC if it has 
been fragmented or altered.

\begin{enumerate}
\setcounter{enumi}{\thenumberedCntCC}
\item Storage efficiency
\setcounter{numberedCntCC}{\theenumi}
\end{enumerate}
mARC auto-regulates the amount of storage allocated to index the 
contextual information. The size of the context information depends on 
the density of the relationships in the data set but is bounded by 
\textit{O (log n)} of the data set size.

For plain text data, the context space typically evolves from O (n) for 
a small data set to O (log n).

\begin{enumerate}
\setcounter{enumi}{\thenumberedCntCC}
\item Ease of programming
\setcounter{numberedCntCC}{\theenumi}
\end{enumerate}
The mARC APIs provide an easy programmatic access to the context 
information. This allows developers to efficiently develop context-aware 
data management applications.

\section{Applications}

mARC has a broad potential for applications. It is particularly well 
suited to big data applications.

\begin{itemize}
\item Keyword-oriented search engines.
\item Context-oriented search engines. Contextual search is to be 
understood as the intuitive meaning of contexts in free form texts. E.g. 
the terms of a request or of an article, are not to be present in the 
result of a user request, or in a similarity process. Contextual text or 
request processing is able to solve ambiguities, and to extract the 
discriminant or low frequency significant information.
\item Contextual meta search engine, to enhance existing search 
facilities
\item Contextual indexation algorithms to enhance existing search 
facilities
\item User request profiling (solving ambiguous requests by user 
context)
\item User profiling (indexing each user by its requests or other 
criterions)
\item Contextual document routing inside a global information system
\item Contextual document matching with a given static ontology
\item Contextual survey of documents flows
\item Contextual similarity matching between documents
\end{itemize}
\section{Experimental Results}

In order to demonstrate some of the benefits of mARC, we have built a 
basic World Wide Web search engine demonstrator using the mARC APIs. We 
use it to study the performance of mARC-based search engines with that 
of a high-performance procedural search engine: Google search.

\section{mARC Search Engine Demonstrator}

The mARC search engine demonstrator provides search features similar to 
Google search: keyword-based queries and auto-completion of search 
queries.

The demonstrator provides additional functionality not currently 
accessible to procedural search engines:

\begin{itemize}
\item Search for contextually-related articles, called \textit{similar 
article} function in the remainder of the paper.
\item Query auto-completion based on pattern association (noisy 
recognition of misspelled queries).
\item Meta-search engine for image retrieval.
\end{itemize}
For the purpose of this study, the demonstrator has been restricted in 
order to be comparable with a keyword-based or N-gram based search 
engine like Google. The full contextual search engine cannot be used in 
this study because it would not easily allow a side by side comparison 
with a procedural search engine like Google, mainly because it does not 
handle keywords in the Google sense.

Data Corpus

The study is performed on both the English and French Wikipedia 
corpuses. For the comparison, the mARC demonstrator indexes 3.5 million 
English articles and 1 million French articles and Google indexes 3.9 
million English articles and 1.4 million articles.

The demonstrator index is built from local snapshots of the Wikipedia 
French and English corpuses taken previously. On the other hand, the 
Google index is kept up to date quasi-real-time. This explains the 
difference in the number of articles indexed. We do believe, however, 
that the difference in the size of the corpuses does not significantly 
affect the conclusions of this study.

\section{Validity of the Study}

The Google architecture is distributed on a very large scale $[$GSA$]$. 
The demonstrator is hosted on an Intel CoreI5-based server running 
Windows 7. This can make performance comparison claims difficult to back 
due to the difference in architectures, raw computing power, size of the 
indices, network latencies, etc. In the following, we provide elements 
to justify the validity of the comparison.

Google sells search appliances which allow deploying the Google search 
engine within an enterprise. The physical servers sold by Google are 
equivalent in specifications to the one used to run the mARC 
demonstrator. More details about the Google Search Appliance can be 
found at $[$GSA$]$.

Google advertises a minimum 50 ms. response time and an average response 
time of less than one second for a corpus of 300000 to 1000000 documents 
for the Google Search Appliance. Pareto's rule gives an approximate 250 
ms. average response time per request.

The user forums for the Google Search Appliance report a lower 
performance of the Google Search Appliance compared to the Internet 
search engine. Google advertises a 250 ms. average response time for its 
Internet search engine.

The choice of Wikipedia for the analysis is also relevant. Google search 
largely favors Wikipedia when returning search results and Wikipedia 
consistently appears in the top five results returned by Google search 
$[$IPS2012$]$. Google search is highly optimized for Wikipedia. 
Therefore, we believe that restricting the comparison between the mARC 
demonstrator and Google search to the Wikipedia corpus does not put 
Google search at a disadvantage.

Another potential objection to the results presented this study is 
scalability. We are comparing the performance a dedicated demonstrator 
to a search engine which handles three to four billion requests per day 
and indexes 30 billion documents.

Given the structure of the World Wide Web and the redundancy rate in 
documents, Google implements a binary tree for the data. With each 
server managing 10$^{8}$ primary documents, the binary tree is 10 
levels deep for 30$^{.}$10$^{9}$ documents. Therefore, each 
request involves a cluster of at most 10 servers, 11 with an http 
front-end server.

In addition, Google optimizes requests by dispatching the request to 
several clusters in parallel. The cluster which has cached the request 
has the shortest response time. We estimate that the number of 
concurrent cluster varies between 1 and 25 depending on the load. This 
gives us an average number of 120 servers participating simultaneously 
to the resolution of a request. Google advocates 250 servers involved in 
the resolution of each request $[$Google2012$]$.

The Google search infrastructure is dimensioned to sustain 4.10$^{9}$ 
requests per day, which is 46300 requests per second. The number of 
servers to ensure a 1 second response time is 46300 x 11 = 509300. The 
number of servers operated by Google is estimated to be around 1.7 
million so the load of a Google search server is therefore comparable to 
the observed load on the mARC demonstrator server.

These considerations lead us to believe that the response time 
comparison between Google search and the mARC demonstrator is valid.

In the following sections, we analyze some of the results gathered with 
the mARC search engine demonstrator to evaluate how the mARC claims 
stand up to experimentation.

\section{Independence from the Data Set}

In the demonstrator, indexation and search are identical for the English 
and French corpuses. There is no language-specific customization. We can 
easily demonstrate the same independence from the data set on the 
Wikipedia corpus in other languages.

However, we have made two simplifying assumptions in the implementation 
of the demonstrator:

\begin{itemize}
\item The input signal is segmented into 8 bits packets.
\item The space character is implicitly used to segment the input 
signal.
\end{itemize}
As a consequence of this simplification, the demonstrator does not 
currently allow the validation of the claim of universal independence 
from the data. Nevertheless, it proves \textit{a minima} the 
independence from the language.

\section{Storage Efficiency}

The following table presents the size in MB of various data elements for 
the mARC search engine demonstrator: size of the mARC contextual RAM, 
size of the index and the inverse resolution database, as well as the 
stored data set size corresponding to the whole English and French 
Wikipedia corpuses.

\begin{table}[h]
\centering
\begin{tabular}{|l|l|l|l|l|l|l|}
\hline
Corpus & mARC & Index & Inverse resolution & Total & Data & Ratio \% \\
\hline
Fr & 500 & 900 & 731 & 2100 & 4000 & 52.5 \\
\hline
En & 600 & 1600 & 1500 & 3700 & 11000 & 33.6 \\
\hline
\end{tabular}
\end{table}

From this data, we observe the following:

\begin{itemize}
\item The size of the mARC does not grow linearly with the size of the 
data set. Rather, it grows in log (data size).
\item The size of the index is at most around 50\% of the size of the 
data set. The index contains all the information necessary to implement 
the search functionality.
\end{itemize}
It should be noted that comparable full text search functionality 
provided by relational database vendors or search engines such as Indri 
or Sphinx requires indices which are 100\% to 300\% of the data set size 
$[$Turtle2012$]$ depending on the settings of the underlying indexation 
API. The size of the Google index was not available at the time of 
writing.

Furthermore, mARC is at a relative disadvantage when doing keyword-based 
search (which is needed for this comparison). A mARC-based search engine 
using the context information more directly (as exemplified by the 
similar article feature of the demonstrator) would leverage more of the 
power of mARC. This approach would reduce the overall memory footprint 
of the mARC search engine metadata by one order of magnitude.

\section{Response Time}

We have measured the response time for the two search engines over two 
classes of requests:

\begin{itemize}
\item Popular queries. A set of a hundred requests among the most 
popular for English and French Wikipedia at the time of the study 
$[$techxav2009$]$.
\item Complex queries. For this measurement, we use the title of a 
Wikipedia article returned by the search engine in response to a query 
as the query (i.e. copy/paste). This allows us to take into account the 
trend towards larger requests which has been observed in recent years 
$[$WIKI2001$]$.
\end{itemize}
In order to account for any caching effects, each query is run four 
times in the experiments. The first time to measure the response time 
for a non-cached request and the subsequent times to average the 
response time after the request has been cached.

We measure the response time for each query run. The response time 
reported in the results only accounts for the wall clock time taken by 
the search engine to resolve the requests. We exclude all network, 
protocol, and response formatting overheads from the analysis.

The average response time is extrapolated using Pareto's 80/20 rule: 

\ \ \ \ First request (non-cached) * 0.2 + average (next 3 requests) * 
0.8

In addition, the true recall rate is also measured. It should be noted 
that for the Google search engine, the recall rate returned by the 
server in the query result is potential. E.g.:

About 158,000,000 results (0.19 seconds)

We measure the true recall rate by navigating to the last page of 
results returned by Google. In order to limit the results to the most 
relevant articles, Google prunes out articles with similar contents. 
Including the pruned-out articles in the query does not significantly 
affect the recall rate. In our measurements, the real recall rate never 
exceeded 800 results.

For the mARC demonstrator, the real recall rate is displayed. All 
articles are directly accessible from the results page.

The experimental results are summarized in the following table:

\begin{table}[h]
\centering
\begin{tabular}{|c|c|c|c|c|c|c|}
\hline
  & \multicolumn{3}{c|}{ Avg. response time (ms.) } & \multicolumn{3}{|c|}{Real recall rate (articles)} \\
\hline 
 & mARC & Google & \textbf{Ratio} & mARC & Google & \textbf{Ratio} \\
\hline
Popular queries (EN) & 12.3 & 132.3 & \textbf{16.4} & 808 & 621 & 
\textbf{1.30} \\
\hline
Popular queries (FR) & 11.6 & 119.1 & \textbf{20.5} & 647 & 415 & 
\textbf{1.56} \\
\hline
Complex queries (EN) & 19.3 & 261.3 & \textbf{15} & 887 & 302 & 
\textbf{2.94} \\
\hline
Complex queries (FR) & 13.3 & 279.2 & \textbf{24.1} & 778 & 299 & 
\textbf{2.60} \\
\hline
\end{tabular}
\end{table}

The results show significantly better response times for the mARC 
demonstrator. The detailed results are presented in appendix 1.

In terms of computing resources, the mARC CPU utilization on the 
demonstrator is measured to less than 10\% of the response time. The 
remainder of the response time is disk access, formatting, API and 
communication overhead. Similar results are not available for Google.

For the popular requests, the average response time for Google when 
restricted to the domains en.wikipedia.org and fr.wikipedia.org are 
respectively 119 and 132 ms. The response time for the same requests 
without the domain restriction is around 320 ms. This measurement is 
consistent with Google's advertised average response time of 250ms. From 
this we can deduce that:

\begin{itemize}
\item Google optimizes the response time for popular domains, such as 
Wikipedia.
\item The Google servers are lightly loaded, as indicated by the small 
variance of response times.
\end{itemize}
This gives us reasonable confidence that the results reported in this 
paper are meaningful. mARC shows response times over an order of 
magnitude better than Google (see Appendix for numerical details).

It should be noted that with mARC once the initial results page has been 
access, all the results have been cached. As a consequence, the average 
access time to a page containing the next 20 results is in the order of 
5 ms. With Google, displaying the next results is equivalent to issuing 
a new (non-cached) request between 70 and 300 ms. for each page.

In addition, it should be noted that the mARC demonstrator does not 
perform any optimization on the request itself. Each result page change 
causes the request to be completely re-evaluated, as in Google search. A 
trivial optimization would be to keep the results in a session variable 
to optimize the scanning of the cached results on the mARC server. This 
would reduce the response time to about 0.5 ms. per results page, 
independently of the complexity of the query.

Given that in practice the average request generates navigation to 2.5 
pages, we can interpolate the average response time of a mARC-based 
search engine to be less than 5 ms. which is 25 times faster than Google 
search.

\section{Search Relevance}

Even though search relevance is a largely subjective notion and cannot 
be accurately measured, it is nevertheless very real and important. The 
only valid measurement technique would be some form of double blind 
testing rating user satisfaction. Nevertheless, we attempt in the 
following to provide some insights into the differences in relevance 
between a mARC-based search engine and a procedural search engine like 
Google.

The Google search algorithm is well documented in the literature. In the 
following, we focus on describing the search strategies implemented in 
the mARC demonstrator centered around:

\newcounter{numberedCntDG}
\begin{enumerate}
\item Keyword-based queries and more pattern-sensitive detection.
\item Context similarity-based queries and associative-sensitive 
detection. 
\setcounter{numberedCntDG}{\theenumi}
\end{enumerate}
The mARC search demonstrator resolves queries using both search 
strategies in parallel. The results are displayed in two columns on the 
results page to allow for easy comparison as shown in figure 2 below.

\begin{figure}[h]
\centering

\includegraphics[width=16.01cm,height=8.23cm]{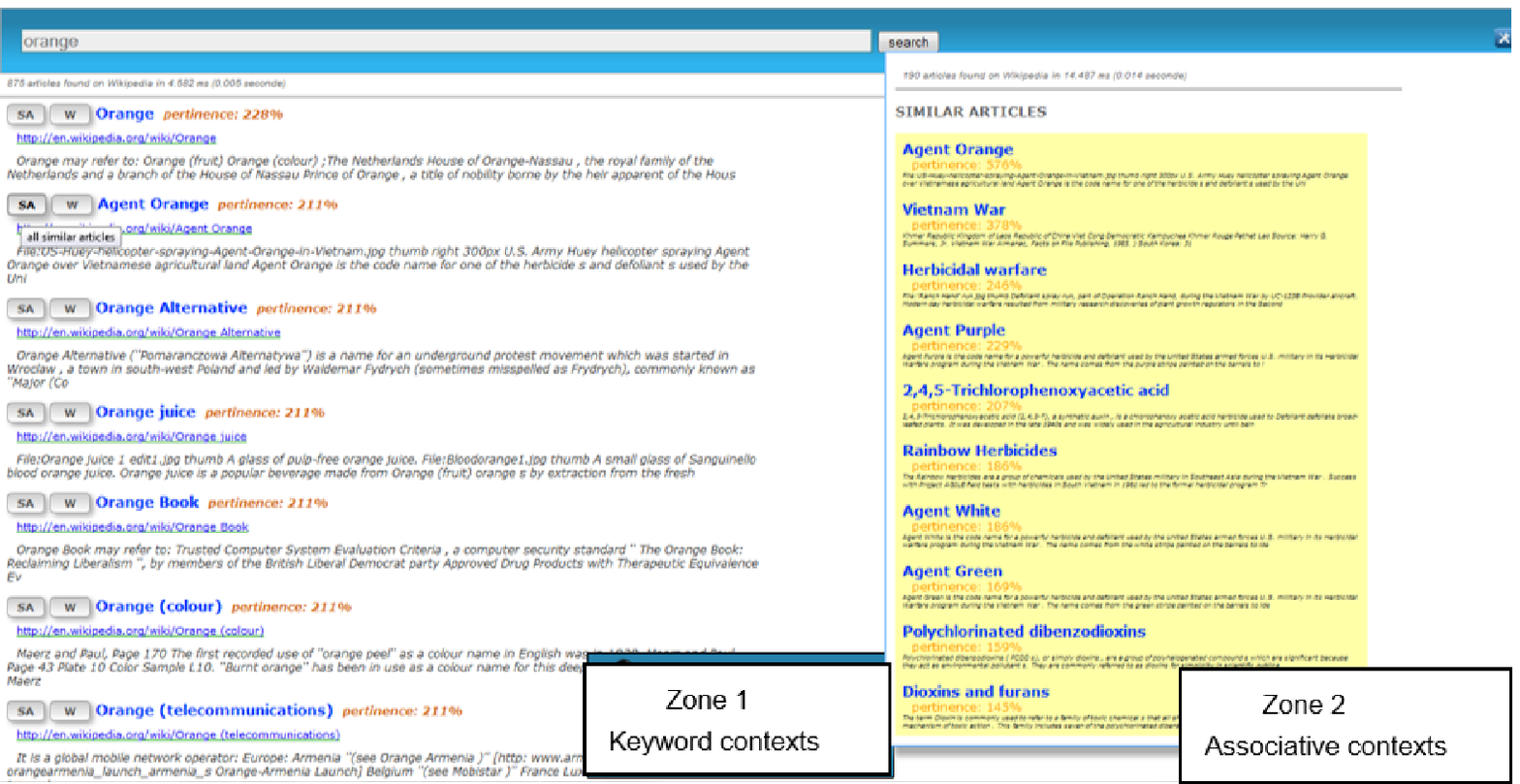}

\includegraphics[width=16.15cm,height=8.17cm]{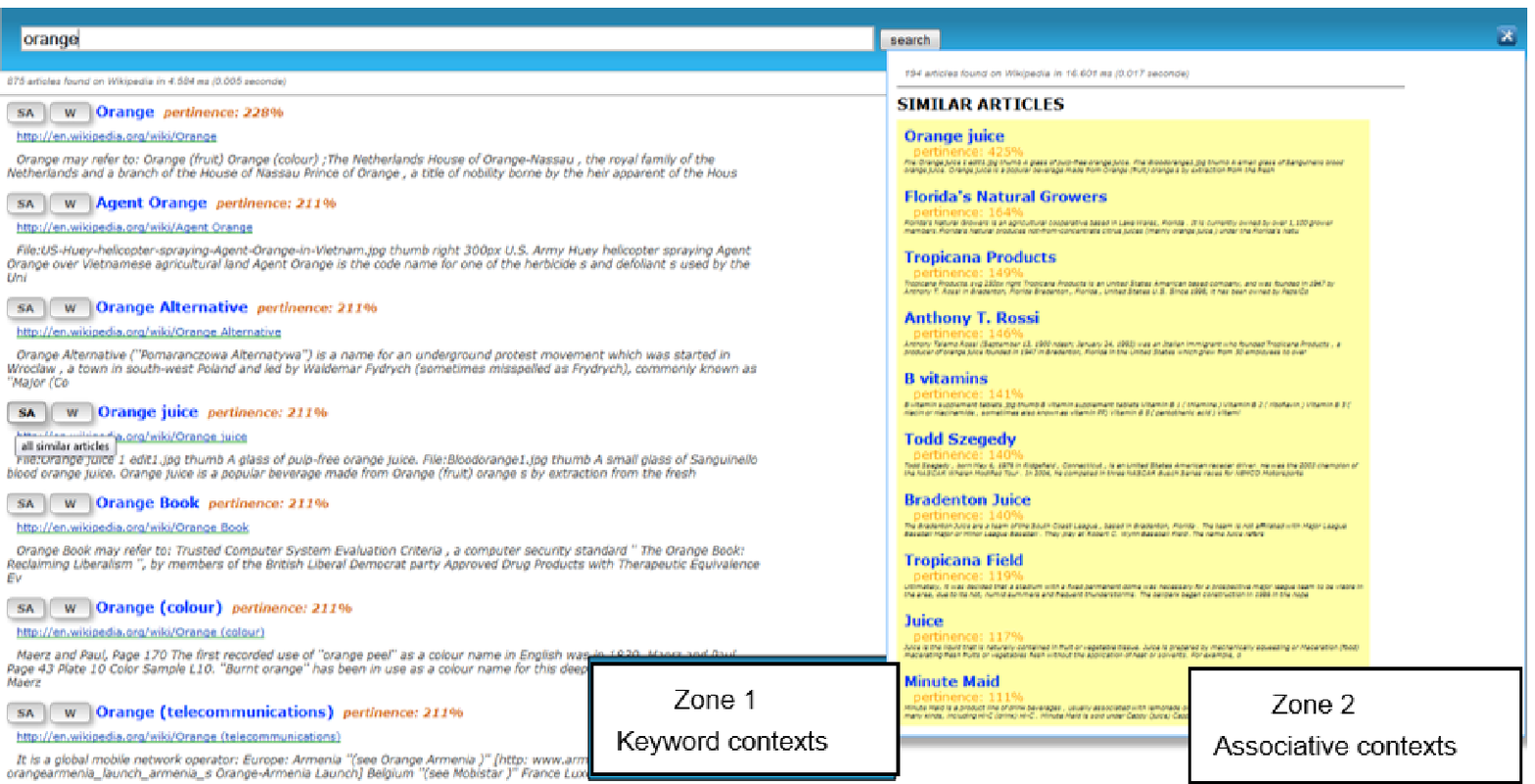}

\caption{Figure 2: mARC demonstrator search results}
\end{figure}

\section{Keyword-based Search}

The first column (Zone 1) displays the results for the keyword-based 
search. The current implementation of mARC does not feature customizable 
indexation strategies. Therefore, keyword-based search is only 
approximated through the API.

In this mode, there is no contextual evaluation of the request. This 
restriction allows the demonstrator to emulate as closely as possible 
the operation of a procedural search engine like Google. To this effect, 
the query routine favors elementary word contexts over associative 
contexts. In practice, the behavior is essentially similar to a pure 
keyword-based approach, nevertheless with a touch of implicit 
associativity.

We observe the following trends:

\begin{itemize}
\item For generic and well know query terms, the shape is preponderant. 
Both the titles and the bodies of the Wikipedia articles returned as 
results contain the keywords. Compared to a pure keyword-based request 
which only returns normalized relevance with respect to the matched 
keywords, the contextual activation greater than 100\% indicates that 
the returned article is contextually over-activated and thus 
contextually plainly meaningful. In other words, the non-normalized (as 
in this demonstrator) confidence rate over 100\% means that the 
resulting documents contain not only the keywords, as a significant 
pattern, but also a part of the contexts associated with this keyword 
inside the knowledge.
\end{itemize}
Example query: \textit{programming}.

\begin{itemize}
\item For more qualified queries, associativity becomes preponderant. 
This means that the articles ranked as most relevant by mARC may not 
contain the terms of the request.
\end{itemize}
Example query: \textit{Ferdinand de Saussure}

Out of the 41 first results returned (all evaluated to be relevant by a 
panel of human observers), 9 results have been retrieved through 
associative contexts and do not contain the terms Ferdinand de Saussure.

There are more differences in behavior compared to a purely procedural 
approach.

Adding terms to a query is equivalent to adding shape contexts. The 
contexts interact with each other. Since the search is focused on shape, 
this is equivalent to an intersection of keywords for the most activated 
articles. The search results are ordered by the analogy with the shape 
of the request (in the title or in the body of the article). Then, as 
the activation decreases, sub-contexts start appearing up to a point of 
disjoint shape sub-contexts with low activation. In this configuration, 
activation $<$=100 implies a shape and association more and more 
disjoint from the request.

Article titles are not privileged. But since titles are small contexts, 
the relative importance of each term is more contrasted.

Example queries: \textit{thallium, skirt, oxygen, oxygen + nitrogen, 
octane rating}

In summary, for shape-based search we observe the following when 
comparing the results returned by the mARC demonstrator and Google:

\begin{itemize}
\item One term queries
\end{itemize}
Example queries: \textit{history, orange, metempsychosis}

For generic requests like \textit{history}, the results are very 
similar; only differing in the order of the results. The mARC 
demonstrator has a slight tendency to order the results in categories: 
history and geographical context (history of various countries), then 
history of well know countries (U.S.A., France, etc.), then the broad 
general categories such as history of sciences, history of literature, 
economy, military, etc.

For generic and ambiguous requests like \textit{orange}, the behavior 
is roughly the same for both search engines. However, we observe a 
slightly better tendency for mARC to vary the semantic contexts on the 
first few result pages.

For targeted requests (i.e. request that do not yield a lot of results), 
we observe that the mARC demonstrator returns significantly more 
relevant results than Google. The reason is that semantic contexts are 
weighted more heavily and return matches for both form and substance. 
The return rate is higher.

Overall, we observe that the more targeted the request, the more 
relevant the results returned by the mARC demonstrator are.

\begin{itemize}
\item Two or three terms queries
\end{itemize}
We observe that the two search engine can return very different results 
for these request:

\begin{itemize}
\item If the terms have little relationship between them (e.g. \textit{
vertebrate politics}), Google returns a list of articles containing 
all terms but without real semantic connection. To the contrary, the 
mARC demonstrator tries to consolidate the two contexts and varies the 
results on the first few result pages. Articles containing all terms are 
generally not activated enough to be presented.
\item If the terms are connected with equivalent generalities, both 
search engines return comparable results, e.g. for \textit{Roddenberry 
}and \textit{Spock}.
\item If the terms are connected with disparate generalities, e.g. 
\textit{wine} and \textit{quantum} Google returns more relevant 
results. The mARC demonstrator tends to return only one to five 
seemingly relevant articles.
\item If the terms are precise, the mARC context associativity kicks in. 
More results are returned and are more relevant than Google's (e.g. 
\textit{Stegastes fuscus, Tantalum 180m, Niobe daughter Tantalus, 
Amyclas, hemoprotein, cyanide intoxication, organophosphate 
intoxication, chrome cancer (professional disease), anaerobic 
respiration}).
\end{itemize}
It should be noted that Google is not very sensitive to the ordering of 
terms within the query. The mARC demonstrator can be if the order 
carries a semantic change. E.g. \textit{red green} and \textit{green 
red} are treated as equal by the mARC while \textit{Paris Hilton} and 
\textit{Hilton Paris} are not.

Overall, we find that Google search provides slightly more relevant 
results in the case of keywords-based search. This can be easily 
understood. On one hand Google search relies intensively rely on user 
requests to improve the search results. On the other hand we, as humans, 
use Google search on a daily basis. In a way we are self-trained to know 
what results to expect. It is in this query range that the trio 
intersection/return rate/relevance is the least random.

However, the real-time article similarity matching provided by the mARC 
demonstrator offers dynamic query disambiguation capabilities which are 
out of the reach of Google search.

\begin{itemize}
\item Long queries
\end{itemize}
These requests are either article titles (four or more terms) or 
copied/pasted from article text.

For these queries, the mARC demonstrator provides indisputably more 
relevant, better categorized results than Google search. The request 
contains enough contextual information for the mARC to evaluate and 
classify the articles in a more relevant way than Google search.

On a number of categorical articles from Wikipedia, we observe that the 
mARC demonstrator and Google search return very similar results for the 
first two or three results pages.

Example: \textit{list of IATA airport codes}

Surprisingly, Google search returns the correct article as result, even 
though the article does not contain all of the terms of the query. The 
reason for this behavior is that the Wikipedia article files often 
contain links (added by the authors) which point to articles relevant to 
the topic. Google uses these links to improve its search relevance. The 
mARC does not use this metadata and only considers the text of the 
articles.

It is interesting to note that both search engines return the same 
results in this case. This emphasizes the ability of the mARC to detect 
semantic relationships. mARC does a comparable job in finding semantic 
relationships between articles as the Wikipedia authors.

\section{Similar Article Search}

The similar article search results are displayed on the right column on 
the results page (Zone 2). A similar capability is not available for 
Google search.

Example: \textit{Orange} and \textit{SA} ("Similar Articles" search 
button) in the different articles returned in Zone 1.

 We find that the similar article search feature enhances the 
keyword-based search results in a very interesting and significant way. 
The conjunct use of the pattern-based search and similarity-based search 
allows a semantic-driven navigation from the initial query with low risk 
of ambiguity. It gives access to different, yet relevant, results which 
are not accessible through keyword-based search.

In addition, similarity-based search helps categorize the results of the 
keyword-based query.

This is a novel and unique feature of mARC.

\section{Ease of Programming}

The PHP code which implements the similar article functionality in the 
demonstrator is shown below. The context detection and selection logic 
is entirely provided in a generic manner by mARC.

public function connexearticles (\$rowid) \{

// similar article

\$this-$>$s-$>$Execute(\$this-$>$session, 
'CONTEXTS.CLEAR');\\
\$this-$>$s-$>$Execute(\$this-$>$session, 
'RESULTS.CLEAR');\\
\$this-$>$s-$>$Execute(\$this-$>$session, 
'CONTEXTS.SET','KNOWLEDGE',\$this-$>$knw );\\
\$this-$>$s-$>$Execute(\$this-$>$session, 
'CONTEXTS.NEW');\\
\$this-$>$s-$>$Execute(\$this-$>$session, 
'TABLE:wikimaster2.TOCONTEXT',\$rowid);\\
\$this-$>$s-$>$Execute(\$this-$>$session, 
'CONTEXTS.DUP');\\
\$this-$>$s-$>$Execute(\$this-$>$session, 
'CONTEXTS.EVALUATE');\\
\$this-$>$s-$>$Execute(\$this-$>$session, 
'CONTEXTS.FILTERACT','25','true');\\
\$this-$>$s-$>$Execute(\$this-$>$session, 
'CONTEXTS.NEWFROMSEM','1','-1','-1');\\
\$this-$>$s-$>$Execute(\$this-$>$session, 
'CONTEXTS.SWAP');\\
\$this-$>$s-$>$Execute(\$this-$>$session, 
'CONTEXTS.DROP');\\
\$this-$>$s-$>$Execute(\$this-$>$session, 
'CONTEXTS.SWAP');\\
\$this-$>$s-$>$Execute(\$this-$>$session, 
'CONTEXTS.DUP');\\
\$this-$>$s-$>$Execute(\$this-$>$session, 
'CONTEXTS.ROLLDOWN','3'); \\
\$this-$>$s-$>$Execute(\$this-$>$session, 
'CONTEXTS.UNION');\\
\$this-$>$s-$>$Execute(\$this-$>$session, 
'CONTEXTS.EVALUATE');\\
\$this-$>$s-$>$Execute(\$this-$>$session, 
'CONTEXTS.INTERSECTION');\\
\$this-$>$s-$>$Execute(\$this-$>$session, 
'CONTEXTS.NORMALIZE');\\
\$this-$>$s-$>$Execute(\$this-$>$session, 
'CONTEXTS.FILTERACT','25','true' );\\
\$this-$>$s-$>$Execute(\$this-$>$session, 
'CONTEXTS.TORESULTS','false','25');\\
\$this-$>$s-$>$Execute(\$this-$>$session, 
'RESULTS.SelectBy','Act','$>$','95');\\
\$this-$>$s-$>$Execute(\$this-$>$session, 
'RESULTS.SortBy','Act','false');\\
\$this-$>$s-$>$Execute(\$this-$>$session, 
'RESULTS.GET','ResultCount');\\
\$count = \$this-$>$s-$>$KMResults;\\
\}

\section{Conclusion and Future Work}

This paper has presented the basic principles of mARC and studied its 
application to Internet search. The results indicate that a mARC-based 
search engine has the potential to be an order of magnitude faster yet 
more relevant than current commercial search engines.

In the current mARC implementation, sampling of the incoming signal is 
limited to eight bits. We are currently working on improving the 
sensorial layer (reading head) to sample UTF-8 signals. This will enable 
the mARC search engine to read and learn complex scripted languages such 
a Chinese, Vietnamese, Hindi or Arabic and all other languages.

In a later stage, we will investigate non-sampled incoming signals to 
enable mARC to process any kind of noisy, weakly-correlated signal.

Finally, we are also working on other application domains of the mARC 
besides text mining. 

\section*{Acknowledgements}

The authors would like to thank Prof. Claude Berrou, member of the 
French Academy of sciences and IEEE Fellow for his comments on earlier 
revisions of this document.

\section*{References}

$[$ACE$]$ Automatic content extraction (ACE) evaluation corpus.\\
http://www.itl.nist.gov/iad/mig/tests/ace.

$[$Hockenmaier$]$ Julia Hockenmaier, "\,Hadoop/Mapreduce in NLP and 
machine learning", Siebel Center.

$[$WIKI01$]$ http://en.wikipedia.org/wiki/Web\_search\_query

$[$Einstein1935$]$ A. Einstein, B. Podolsky and N. Rosen, Can 
quantum-mechanical description of physical reality be considered 
complete ?, Phys. Rev., vol.47, 1935.

$[$Papert1969$]$ $[$Papert1969$]$ S. Papert and M.Minsky, Perceptrons, 
An Introduction to Computational Geometry, MIT press, Cambridge, 
Massassuchets, 1969.

 $[$Clark1973$]$ The language-as-fixed-effect fallacy: A critique of 
language statistics in psychological research, Journal of verbal 
learning and verbal behavior, Elsevier, 1973.

$[$Aspect1981$]$ A. Aspect, P. Grangier, G. Roger, Phys. Rev. Lett., 47, 
460 (1981). 

$[$Aspect1982$]$ A. Aspect, P. Grangier, G. Roger, Phys. Rev. Lett., 49, 
91 (1982).

$[$Penrose1989$]$ R. Penrose. The Emperor's New Mind (Oxford, Oxford 
Univ. Press, 1989).

$[$Kupiec1992$]$ J. Kupiec. Robust part-of-speech tagging using a hidden 
Markov model. Computer Speech \& Language, 6(3):225-242, 1992.

$[$Grishman1993$]$ Ralph Grishman and Beth Sundheim. Message 
understanding conference-6: A brief history. In Proceedings of the 16th 
International Conference on Computational Linguistics, pages 466-471, 
1996.

$[$Penrose1997$]$ R. Penrose, in the Large, the Small and the Human 
Mind, ed. M. Longair (Cambridge, Cambridge Univ. Press, 1997)

$[$Nigam1998$]$ K. Nigam, A. McCallum, S. Thrun, T. Mitchell. Learning 
to classify text from labeled and unlabeled documents, AAAI Conference, 
1998.

$[$Bikel1999$]$ D. Bikel, R. Schwartz, and R. Weischedel, An algorithm 
that learns what's in a name. Machine learning, 34(1):211-231, 1999.

$[$GSA$]$ Google Search Appliance. http://support.google.com/gsa

$[$Hofmann1999$]$ T. Hofmann. Probabilistic Latent Semantic Indexing. 
ACM SIGIR Conference, 1999.

$[$Rosenfeld2000$]$ R Rosenfeld, Two decades of statistical language 
modeling: Where do we go from here? Proceedings of the IEEE, 2000.

$[$Lafferty2001$]$ J. D. La?erty, A. McCallum, and F. C. N. Pereira. 
Conditional random ?elds: Probabilistic models for segmenting and 
labeling sequence data. In ICML, pages 282-289, 2001.

$[$Sarwar2001$]$ Sarwar, B. M., Karypis, G., Konstan, J. A., and Riedl, 
J. "Item-based Collaborative Filtering Recommendation Algorithms". ACM 
WWW10 Conference, May, 2001.

$[$Basu2002$]$ S. Basu, A. Banerjee, R. J. Mooney. Semi-supervised 
Clustering by Seeding. ICML Conference, 2002.

$[$Burke2002$]$ Robin Burke, User Modeling and User-Adapted Interaction 
Volume 12 Issue 4, Pages 331 - 370 November 2002. 

$[$Burke2002-2$]$ Hybrid Recommender Systems: Survey and Experiments, 
Robin Burke, User Modeling and User-Adapted Interaction, Volume 12, 
Issue 4, pp. 331-370, November 2002.

$[$Andrieu2003$]$ C. Andrieu, N. DeFreitas, A. Doucet and M. Jordan. An 
introduction to mcmc for machine learning. Machine learning, 50(1):5-43, 
2003.

$[$Bengio2003$]$ Bengio, Y., R. Ducharme, P. Vincent, and C. Jauvin. A 
neural probabilistic language model. Journal of Machine Learning 
Research, 3:1137-1155, 2003.

$[$Blei2003$]$ D. Blei, A. Ng, M. Jordan. Latent Dirichlet allocation, 
Journal of Machine Learning Research, 3: pp. 993-1022, 2003.

$[$McCallum2003$]$ A. McCallum and W. Li. Early results for named entity 
recognition with conditional random ?elds, feature induction and 
web-enhanced lexicons In Proceedings of the seventh conference on 
Natural language learning at HLT-NAACL 2003-Volume 4, pages 188-191. 
Association for Computational Linguistics, 2003.

$[$Sha2003$]$ F. Sha and F. Pereira. Shallow parsing with conditional 
random ?elds In Proceedings of the 2003 Conference of the North American 
Chapter of the Association for Computational Linguistics on Human 
Language Technology-Volume 1, pages 134-141, 2003.

$[$Tjong2003$]$ Erik F. Tjong Kim Sang and Fien De Meulder. Introduction 
to the CoNLL-2003 shared task: Language-independent named entity 
recognition. In Proceedings of the 7th Conference on Natural Language 
Learning, pages 142-147, 2003.

$[$Basu2004$]$ S. Basu, M. Bilenko, R. J. Mooney. A probabilistic 
framework for semi-supervised clustering. ACM KDD Conference, 2004.

$[$Borman2004$]$ Borman S., The Expectation Maximization Algorithm -- A 
short tutorial.

$[$Dean2004$]$ Dean, Jeffrey \& Ghemawat, Sanjay "MapReduce: Simplified 
Data Processing on Large Clusters", 2004.

$[$Rijsbergen2004$]$ The geometry of Information Retrieval, Cambridge 
University Press, 2004.

$[$USP2004$]$ Method and system for adaptive learning and pattern 
recognition, US Patent Application No: 2004/0205, 035. 

$[$Saragawi2005$]$ Sunita Sarawagi and William W. Cohen. Semi-Markov 
conditional random ?elds for information extraction. In Advances in 
Neural Information Processing Systems 17, pages 1185-1192. 2005.

$[$Jiang2006$]$ Jing Jiang and ChengXiang Zhai. Exploiting domain 
structure for named entity recognition. In Proceedings of the Human 
Language, Technology Conference of the North American Chapter of the 
Association for Computational Linguistics, pages 74-81, 2006.

$[$Gupta2007$]$ S. Gupta, A. Nenkova and D. Jurafsky. Measuring 
importance and\\
query relevance in topic-focused multi-document summarization In 
Proceedings of the Annual Meeting of the Association for Computational 
Linguistics, Demo and Poster Sessions, pages 193-196, 2007.

$[$Banko2008$]$ Michele Banko and Oren Etzioni. The tradeo?s between 
open and traditional relation extraction. In Proceedings of the 46th 
Annual Meeting of the Association for Computational Linguistics, pages 
28-36, 2008.

$[$LongHua2008$]$ Longhua Qian, Guodong Zhou, Fang Kong, Qiaoming Zhu, 
and Peide Qian. Exploiting constituent dependencies for tree kernel 
based semantic relation extraction \textit{In} Proceedings of the 22nd 
International Conference on Computational Linguistics, pages 697-704, 
2008.

$[$techxav2009$]$ http://www.techxav.com/2009/08/31/wikipedia/

$[$Almazro2010$]$ Dhoha Almazro, Ghadeer Shahatah, Lamia Albdulkarim, 
Mona Kherees, Romy Martinez, William Nzoukou, A Survey Paper on 
Recommender Systems, Cornell University Library. $[$Lok2010$]$ Corie Lok 
"Speed Reading", NATURE, vol. 463, January 2010.

$[$Etzioni2011$]$ Oren Etzioni, "Search needs a shake-up", Nature, vol. 
476 August 2011.

$[$Foto2012$]$ Foto N. Afrati and Anish Das Sarma and Semih Salihoglu 
and Jeffrey D. Ullman, Vision Paper: Towards an Understanding of the 
Limits of Map-Reduce Computation. http://arxiv.org/abs/1204.1754.

$[$Aggarwal2012$]$ Mining Text Data. Charu C. Aggarwal, ChengXiang Zhai, 
Springer, February 2012.

$[$DU2012$]$ Dong Yu, Goeffrey Hinton, Nelson Morgan, Jen-Tzung Chien, 
Shigeki Sagayame, IEEE Transactions on Audio, Speech, and Language 
Processing, vol. 20, no. 1, January 2012.

$[$Google2012$]$ "\,Web search for a planet: the Google cluster 
architecture" - \textit{
research.google.com/archive/googlecluster-ieee.pdf}

$[$IPS2012$]$ \underline{
http://www.intelligentpositioning.com/blog/2012/02/wikipedia-page-one-of-google-uk-for-99-of-searches/
}

$[$Maden2012$]$ Sam Maden, From Databases to Big Data in Internet 
Computing, IEEE Volume: 16, Issue 3, 2012.

$[$Ming2012$]$ Tan, Ming Zhou, Wenli Zheng, Lei; Wang, Shaojun, A 
Scalable Distributed Syntactic, Semantic, and Lexical Language Model 
\textit{in} Computational Linguistics, MIT Press, 2012.

$[$Misyak2012$]$ JB Misyak, MH Christiansen, Statistical learning and 
language: an individual differences study, Language Learning, Wiley 
Online Library, 2012.

$[$Singh2012$]$ Speech Recognition with Hidden Markov Model: A Review, 
International Journal of Advanced Research in Computer Science and 
Software Engineering, Volume 2, Issue 3, March 2012.

$[$Teyssier2012$]$ M Teyssier, D Koller, Ordering-Based Search: A Simple 
and Effective Algorithm for Learning Bayesian Networks \textit{in }
Proceedings of the Twenty-First Conference on Uncertainty in Artificial 
Intelligence (2012).

$[$Turtle2012$]$ H. Turtle, \textit{
opensearchlab.otago.ac.nz/paper\_12.pdf}

$[$Zhou2012$]$ The state-of-the-art in personalized recommender systems 
for social networking, X Zhou, Y Xu, Y Li, A Josang, C Cox - Artificial 
Intelligence Review, 2012 - Springer

$[$Pu2012$]$ Evaluating recommender systems from the user's perspective: 
survey of the state of the art, Pearl Pu, Li Chen, Rong Hu, User 
Modeling and User-Adapted Interaction, Volume 22, Issue 4-5, pp. 317-355 
(2012).

\section{Appendix: Experimental Results}

\begin{table}[t]
\begin{tabular}{|p{2cm}|*{13}{p{0.5cm}|}}
\hline
Wikipedia En & \multicolumn{6}{p{3cm}|}{mARC} & \multicolumn{1}{p{0.5cm}|}{\textbf{Ratio}} & \multicolumn{6}{p{3cm}|}{Google} \\
\hline
 & & & \\
\hline
 & Returned Results (real) & Req1 (ms.) out of cache & Req 2 (ms.) & Req3 
(ms.) & Req 3 (ms.) & average (ms.) & ratio & Returned Results (real) & 
Req1 (ms.) out of cache & Req 2 (ms.) & Req3 (ms.) & Req 3 (ms.) & 
average (ms.) \\
\hline
2012 & 800 & 17 & 5.2 & 5.3 & 5.26 & 7.6 & 20.5 & 800 & 220 & 160 & 130 
& 130 & 156.0 \\
\hline
2009 Swine Flu outbreak & 899 & 21 & 5.1 & 5.1 & 4.8 & 8.2 & 18.0 & 531 
& 300 & 100 & 100 & 130 & 148.0 \\
\hline
Abraham Lincoln & 1071 & 23.3 & 5.3 & 5.3 & 5.4 & 8.9 & 14.6 & 667 & 200 
& 100 & 130 & 110 & 130.7 \\
\hline
Adolf Hitler & 1015 & 14.8 & 4.7 & 4.7 & 4.7 & 6.7 & 20.1 & 629 & 250 & 
90 & 130 & 100 & 135.3 \\
\hline
America's Next Top Model & 667 & 13.5 & 3.5 & 3.5 & 3.5 & 5.5 & 23.4 & 
643 & 230 & 110 & 100 & 100 & 128.7 \\
\hline
American Idol & 800 & 19.6 & 5.5 & 5.4 & 5.4 & 8.3 & 17.6 & 659 & 300 & 
90 & 110 & 120 & 145.3 \\
\hline
Anal sex & 962 & 123 & 5.8 & 7.6 & 5.6 & 29.7 & 4.8 & 626 & 260 & 110 & 
130 & 100 & 142.7 \\
\hline
Australia & 651 & 12.3 & 3.7 & 3 & 3.1 & 5.1 & 28.6 & 600 & 300 & 110 & 
100 & 110 & 145.3 \\
\hline
Barack Obama & 1131 & 17.3 & 6.9 & 6.9 & 6.9 & 9.0 & 15.4 & 645 & 250 & 
110 & 120 & 100 & 138.0 \\
\hline
Batman & 683 & 38.6 & 3.5 & 3.5 & 3.4 & 10.5 & 12.7 & 660 & 200 & 140 & 
100 & 110 & 133.3 \\
\hline
Bleach manga & 804 & 30.2 & 6 & 6 & 5.9 & 10.8 & 14.1 & 598 & 280 & 110 
& 140 & 110 & 152.0 \\
\hline
Canada & 800 & 14.2 & 5 & 5 & 5 & 6.8 & 22.4 & 632 & 260 & 120 & 120 & 
140 & 153.3 \\
\hline
China & 989 & 19.3 & 5.9 & 6.3 & 5.9 & 8.7 & 19.4 & 600 & 270 & 140 & 
160 & 130 & 168.7 \\
\hline
Current events portal & 897 & 254 & 4.4 & 4 & 4 & 54.1 & 2.3 & 665 & 230 
& 90 & 100 & 110 & 126.0 \\
\hline
Deadpool comics & 281 & 12 & 1.5 & 1.5 & 1.4 & 3.6 & 42.5 & 596 & 320 & 
100 & 110 & 120 & 152.0 \\
\hline
Deaths in 2009 & 800 & 18.71 & 8.6 & 5 & 5 & 8.7 & 16.2 & 700 & 250 & 
110 & 100 & 130 & 140.7 \\
\hline
Facebook & 800 & 12.9 & 5.3 & 5.4 & 5.3 & 6.8 & 19.3 & 641 & 220 & 100 & 
120 & 110 & 132.0 \\
\hline
Family Guy & 800 & 11.9 & 5.5 & 5.5 & 5.5 & 6.8 & 24.5 & 676 & 190 & 220 
& 110 & 150 & 166.0 \\
\hline
Farrah Fawcett & 875 & 11 & 5.4 & 5.1 & 5.2 & 6.4 & 23.5 & 502 & 310 & 
110 & 110 & 110 & 150.0 \\
\hline
Favicon.ico & 144 & 29.12 & 0.9 & 0.9 & 0.9 & 6.5 & 20.2 & 295 & 260 & 
100 & 110 & 90 & 132.0 \\
\hline
Featured content portal & 1220 & 8.9 & 7.9 & 7.9 & 7.9 & 8.1 & 14.8 & 
690 & 200 & 100 & 110 & 90 & 120.0 \\
\hline
France & 601 & 22.2 & 2.7 & 2.7 & 2.6 & 6.6 & 23.7 & 700 & 260 & 120 & 
120 & 150 & 156.0 \\
\hline
George W. Bush & 1084 & 107.89 & 6.9 & 6.4 & 6.4 & 26.8 & 4.8 & 646 & 
210 & 110 & 110 & 110 & 130.0 \\
\hline
Germany & 642 & 24.9 & 2.9 & 2.9 & 3 & 7.3 & 20.1 & 700 & 230 & 120 & 
140 & 120 & 147.3 \\
\hline
Global warming & 938 & 7.9 & 4.8 & 4.9 & 4.9 & 5.5 & 27.3 & 632 & 240 & 
150 & 110 & 120 & 149.3 \\
\hline
Google & 800 & 7.9 & 5.9 & 5.5 & 5.5 & 6.1 & 20.0 & 663 & 210 & 90 & 110 
& 100 & 122.0 \\
\hline
Henry VIII of England & 1232 & 70.88 & 10 & 9.5 & 9.6 & 21.9 & 6.4 & 662 
& 260 & 100 & 110 & 120 & 140.0 \\
\hline
Heroes TV series & 800 & 18.5 & 5.6 & 5.7 & 5.6 & 8.2 & 18.3 & 654 & 230 
& 120 & 130 & 140 & 150.0 \\
\hline
Hotmail & 136 & 21.9 & 0.7 & 0.7 & 0.7 & 4.9 & 27.3 & 451 & 180 & 120 & 
110 & 140 & 134.7 \\
\hline
House TV series & 800 & 15.4 & 5.2 & 5.3 & 5.3 & 7.3 & 23.6 & 670 & 340 
& 160 & 110 & 120 & 172.0 \\
\hline
Human penis size & 671 & 24.4 & 3.5 & 3.3 & 3.3 & 7.6 & 18.4 & 499 & 270 
& 100 & 110 & 110 & 139.3 \\
\hline
India & 653 & 13.3 & 3 & 2.9 & 3 & 5.0 & 25.7 & 582 & 220 & 100 & 110 & 
110 & 129.3 \\
\hline
Internet Movie Database & 1126 & 60.66 & 6 & 5.9 & 5.8 & 16.9 & 8.9 & 
669 & 200 & 160 & 130 & 120 & 149.3 \\
\hline
Jade Goody & 1317 & 388.7 & 6.6 & 6.2 & 6.1 & 82.8 & 1.5 & 518 & 250 & 
100 & 100 & 90 & 127.3 \\
\hline
Japan & 800 & 15.8 & 5.2 & 5.1 & 5.1 & 7.3 & 20.1 & 700 & 250 & 120 & 
110 & 130 & 146.0 \\
\hline
Jonas Brothers & 917 & 28.3 & 4.7 & 5.2 & 4.5 & 9.5 & 13.3 & 648 & 190 & 
110 & 110 & 110 & 126.0 \\
\hline
Kim Kardashian & 478 & 55.67 & 2.7 & 2.4 & 2.5 & 13.2 & 8.9 & 529 & 250 
& 80 & 80 & 90 & 116.7 \\
\hline
Kristen Stewart & 1283 & 297 & 6.3 & 6.1 & 6.2 & 64.4 & 2.1 & 593 & 260 
& 90 & 90 & 120 & 132.0 \\
\hline
Lady Gaga & 887 & 22.3 & 4.4 & 4.5 & 4.3 & 8.0 & 14.7 & 640 & 240 & 80 & 
90 & 90 & 117.3 \\
\hline
Lil Wayne & 989 & 41.4 & 4.2 & 4.2 & 4.2 & 11.6 & 9.8 & 644 & 210 & 90 & 
90 & 90 & 114.0 \\
\hline
List of Family Guy episodes & 638 & 14.5 & 3.3 & 3.1 & 3.1 & 5.4 & 21.7 
& 595 & 230 & 90 & 90 & 90 & 118.0 \\
\hline
List of Heroes episodes & 804 & 96.45 & 3.3 & 3 & 3 & 21.8 & 5.2 & 589 & 
210 & 80 & 80 & 110 & 114.0 \\
\hline
List of House episodes & 638 & 13.3 & 3.3 & 3.1 & 3.5 & 5.3 & 21.3 & 626 
& 230 & 90 & 80 & 80 & 112.7 \\
\hline
List of Presidents of the United States & 1152 & 33.2 & 10 & 9.9 & 9.9 & 
14.6 & 9.3 & 700 & 280 & 90 & 110 & 100 & 136.0 \\
\hline
List of sex positions & 1185 & 105.2 & 5 & 4.9 & 5.3 & 25.1 & 5.1 & 588 
& 270 & 100 & 100 & 80 & 128.7 \\
\hline
Lost season 5 & 800 & 5.8 & 5.7 & 5.7 & 5.8 & 5.7 & 21.7 & 654 & 250 & 
90 & 100 & 90 & 124.7 \\
\hline
Martin Luther King Jr & 1293 & 40.3 & 7.9 & 7.9 & 7.8 & 14.4 & 11.8 & 
653 & 380 & 130 & 120 & 100 & 169.3 \\
\hline
Masturbation & 466 & 7.8 & 2.7 & 2.6 & 2.6 & 3.7 & 30.7 & 619 & 190 & 90 
& 80 & 110 & 112.7 \\
\hline
Megan Fox & 1022 & 21.4 & 8.6 & 8.7 & 8.7 & 11.2 & 14.6 & 572 & 270 & 80 
& 120 & 210 & 163.3 \\
\hline
Metallica & 604 & 35.28 & 3.2 & 3.2 & 3.2 & 9.6 & 13.3 & 675 & 240 & 100 
& 90 & 110 & 128.0 \\
\hline
Mexico & 629 & 16.6 & 3 & 3 & 2.9 & 5.7 & 25.2 & 626 & 250 & 120 & 130 & 
100 & 143.3 \\
\hline
Michael Jackson & 943 & 28.57 & 4.4 & 4.3 & 4.3 & 9.2 & 12.9 & 651 & 220 
& 80 & 100 & 100 & 118.7 \\
\hline
Mickey Rourke & 697 & 30.6 & 3.5 & 3.5 & 3.4 & 8.9 & 14.5 & 587 & 260 & 
100 & 90 & 100 & 129.3 \\
\hline
Miley Cyrus & 802 & 26.75 & 4.1 & 4 & 4 & 8.6 & 15.3 & 642 & 230 & 110 & 
100 & 110 & 131.3 \\
\hline
MySpace & 800 & 25.2 & 5.5 & 5.5 & 5.4 & 9.4 & 13.5 & 681 & 210 & 140 & 
90 & 90 & 127.3 \\
\hline
Naruto & 726 & 18.5 & 4.6 & 4.5 & 4.6 & 7.4 & 16.0 & 618 & 240 & 90 & 80 
& 90 & 117.3 \\
\hline
Natasha Richardson & 1449 & 485.6 & 6.9 & 7.1 & 6.8 & 102.7 & 1.3 & 579 
& 190 & 120 & 110 & 120 & 131.3 \\
\hline
New York City & 623 & 29.6 & 3.4 & 3.2 & 3.2 & 8.5 & 22.4 & 700 & 370 & 
150 & 160 & 130 & 191.3 \\
\hline
Penis & 800 & 5.6 & 5.2 & 5.2 & 5.2 & 5.3 & 21.6 & 671 & 210 & 90 & 90 & 
90 & 114.0 \\
\hline
Pornography & 800 & 5.3 & 5.3 & 5.3 & 5.3 & 5.3 & 22.6 & 730 & 240 & 100 
& 80 & 90 & 120.0 \\
\hline
Relapse album & 774 & 20.54 & 5.1 & 5.2 & 5.3 & 8.3 & 13.6 & 601 & 230 & 
80 & 90 & 80 & 112.7 \\
\hline
Rhianna & 375 & 129.8 & 1.4 & 1.3 & 1.3 & 27.0 & 6.0 & 247 & 300 & 140 & 
110 & 130 & 161.3 \\
\hline
Robert Pattinson & 251 & 16.22 & 1.5 & 1.4 & 1.3 & 4.4 & 33.8 & 566 & 
230 & 100 & 100 & 180 & 147.3 \\
\hline
Russia & 642 & 38.85 & 3.1 & 3.1 & 3.1 & 10.3 & 12.6 & 700 & 220 & 110 & 
90 & 120 & 129.3 \\
\hline
Scrubs TV series & 800 & 19.34 & 4 & 3.9 & 4 & 7.0 & 19.7 & 95 & 360 & 
80 & 90 & 80 & 138.7 \\
\hline
Selena Gomez & 648 & 29.43 & 3.1 & 3 & 3 & 8.3 & 14.4 & 618 & 210 & 90 & 
110 & 90 & 119.3 \\
\hline
Sex & 800 & 5.7 & 54 & 5.8 & 5.4 & 18.5 & 7.0 & 636 & 210 & 100 & 140 & 
90 & 130.0 \\
\hline
Sexual intercourse & 961 & 20.22 & 4.8 & 4.8 & 4.8 & 7.9 & 15.8 & 646 & 
250 & 100 & 90 & 90 & 124.7 \\
\hline
Slumdog Millionaire & 550 & 21 & 2.7 & 2.7 & 2.8 & 6.4 & 17.4 & 609 & 
210 & 80 & 90 & 90 & 111.3 \\
\hline
Israel & 663 & 22.34 & 3.6 & 3.3 & 3.7 & 7.3 & 17.3 & 700 & 230 & 90 & 
110 & 100 & 126.0 \\
\hline
Star Trek film & 852 & 14.1 & 4.1 & 4.3 & 4.1 & 6.2 & 21.8 & 651 & 270 & 
100 & 90 & 110 & 134.0 \\
\hline
Swine flu & 589 & 12.72 & 2.6 & 2.5 & 2.7 & 4.6 & 25.4 & 591 & 240 & 90 
& 90 & 80 & 117.3 \\
\hline
Taylor Swift & 791 & 48.64 & 3.8 & 3.5 & 3.6 & 12.6 & 9.2 & 635 & 180 & 
90 & 120 & 90 & 116.0 \\
\hline
Terminator Salvation & 520 & 56.91 & 2.3 & 2.3 & 2.3 & 13.2 & 9.3 & 552 
& 240 & 80 & 110 & 90 & 122.7 \\
\hline
The Beatles & 800 & 18.68 & 5.5 & 5.5 & 5.5 & 8.1 & 13.8 & 700 & 230 & 
80 & 80 & 90 & 112.7 \\
\hline
The Dark Knight film & 984 & 25.3 & 5.2 & 5.1 & 5 & 9.1 & 15.2 & 621 & 
240 & 100 & 140 & 100 & 138.7 \\
\hline
The Notorious B.I.G. & 598 & 16.3 & 3.1 & 3.1 & 3 & 5.7 & 21.1 & 642 & 
190 & 100 & 110 & 100 & 120.7 \\
\hline
Transformers 2 & 800 & 48.22 & 5.2 & 5.1 & 5.3 & 13.8 & 9.0 & 644 & 210 
& 100 & 110 & 100 & 124.7 \\
\hline
Transformers: Revenge of the Fallen & 802 & 18.81 & 5 & 4.9 & 4.9 & 7.7 
& 14.7 & 624 & 220 & 80 & 90 & 90 & 113.3 \\
\hline
Tupac Shakur & 635 & 20.67 & 3.1 & 3.2 & 3.1 & 6.6 & 19.3 & 622 & 200 & 
150 & 80 & 100 & 128.0 \\
\hline
Twilight & 800 & 22.73 & 5.5 & 5.4 & 5.5 & 8.9 & 11.7 & 700 & 190 & 80 & 
90 & 80 & 104.7 \\
\hline
Twilight 2008 film & 800 & 7.3 & 5.6 & 5.4 & 5.5 & 5.9 & 20.9 & 610 & 
200 & 110 & 90 & 110 & 122.7 \\
\hline
Twitter & 800 & 18.95 & 5.4 & 5.4 & 5.5 & 8.1 & 15.2 & 634 & 260 & 90 & 
90 & 90 & 124.0 \\
\hline
United Kingdom & 899 & 27.5 & 4.3 & 4.1 & 4.4 & 8.9 & 14.9 & 600 & 250 & 
100 & 110 & 100 & 132.7 \\
\hline
Vagina & 728 & 14.7 & 5.4 & 4.6 & 4.7 & 6.9 & 15.5 & 640 & 160 & 90 & 
100 & 90 & 106.7 \\
\hline
Valentine's Day & 902 & 52.48 & 5.3 & 5.2 & 5.2 & 14.7 & 8.6 & 667 & 260 
& 100 & 90 & 90 & 126.7 \\
\hline
Vietnam War & 1026 & 33.87 & 4.6 & 4.7 & 4.7 & 10.5 & 13.1 & 700 & 250 & 
120 & 120 & 90 & 138.0 \\
\hline
Watchmen film & 494 & 13.6 & 2.8 & 2.8 & 3 & 5.0 & 23.3 & 531 & 210 & 
110 & 90 & 80 & 116.7 \\
\hline
William Shakespeare & 800 & 15.14 & 5.5 & 5.6 & 8.2 & 8.2 & 14.2 & 660 & 
180 & 100 & 100 & 100 & 116.0 \\
\hline
Windows 7 & 800 & 5.74 & 5.6 & 6.3 & 5.4 & 5.8 & 20.8 & 650 & 240 & 90 & 
90 & 90 & 120.0 \\
\hline
Wolverine comics & 905 & 16.6 & 5.7 & 5.7 & 57 & 21.6 & 6.2 & 700 & 240 
& 120 & 90 & 110 & 133.3 \\
\hline
World War I & 640 & 13.19 & 3.1 & 3.5 & 3.3 & 5.3 & 24.8 & 664 & 240 & 
120 & 90 & 100 & 130.7 \\
\hline
World War II & 1040 & 18.03 & 4.7 & 4.6 & 4.7 & 7.3 & 15.5 & 700 & 210 & 
90 & 90 & 90 & 114.0 \\
\hline
X-Men Origins: Wolverine & 1399 & 52.69 & 10.6 & 10.3 & 10.3 & 18.9 & 
6.2 & 624 & 220 & 90 & 90 & 90 & 116.0 \\
\hline
YouTube & 800 & 5.6 & 5.5 & 5.4 & 5.4 & 5.5 & 19.6 & 641 & 190 & 80 & 90 
& 90 & 107.3 \\
\hline
 & & & & & & & & & & & & & \\
\hline
 & & 41.4 & 5.2 & 4.7 & 5.2 & 12.3 & 16.4 & & 239.4 & 105.4 & 104.9 & 
106.1 & 132.3 \\
\hline
\end{tabular}
\end{table}

\newpage

\begin{table}[h]
\centering
\begin{tabular}{|p{2cm}|*{13}{p{0.5cm}|}}
\hline
2010 request sample & \multicolumn{6}{p{3cm}|}{mARC} & \multicolumn{1}{p{0.5cm}|}{\textbf{Ratio}} & \multicolumn{6}{p{3cm}|}{Google} \\
\hline
wikipedia fr & & & \\
\hline
 &Results (real) & Req1 (ms.) out of cache & Req 2 (ms.) & Req3 
(ms.) & Req 3 (ms.) & average (ms.) & ratio &Results (real) & 
Req1 (ms.) out of cache & Req 2 (ms.) & Req3 (ms.) & Req 3 (ms.) & 
average (ms.) \\
\hline
Facebook & 299 & 7.29 & 2.2 & 1.7 & 1.6 & 2.9 & 38.1 & 620 & 170 & 110 & 
90 & 90 & 111.3 \\
\hline
youtube & 482 & 11.7 & 2.8 & 2.8 & 2.8 & 4.6 & 23.6 & 681 & 180 & 90 & 
90 & 90 & 108.0 \\
\hline
jeux & 800 & 6.04 & 5.9 & 5.9 & 5.9 & 5.9 & 20.6 & 700 & 210 & 110 & 100 
& 90 & 122.0 \\
\hline
you & 840 & 15.28 & 5.3 & 5.4 & 5.3 & 7.3 & 15.1 & 695 & 180 & 90 & 100 
& 90 & 110.7 \\
\hline
yahoo & 639 & 4.29 & 4 & 4 & 4 & 4.1 & 27.1 & 642 & 190 & 90 & 90 & 90 & 
110.0 \\
\hline
tv & 800 & 6.5 & 5.7 & 5.7 & 5.7 & 5.9 & 21.2 & 679 & 220 & 100 & 90 & 
110 & 124.0 \\
\hline
orange & 837 & 11.3 & 5.2 & 5.3 & 5.3 & 6.5 & 19.9 & 689 & 230 & 110 & 
100 & 100 & 128.7 \\
\hline
meteo & 183 & 23.63 & 1.4 & 1.2 & 1.3 & 9.2 & 13.6 & 678 & 180 & 110 & 
90 & 100 & 125.3 \\
\hline
le bon coin & 858 & 21.3 & 6.3 & 6.2 & 6.2 & 9.2 & 13.6 & 498 & 240 & 90 
& 110 & 90 & 125.3 \\
\hline
hotmail & 215 & 17.98 & 1.3 & 1.7 & 1.3 & 4.7 & 26.4 & 570 & 200 & 90 & 
120 & 110 & 125.3 \\
\hline
yahoo mail & 802 & 7.36 & 6.9 & 6.5 & 6.5 & 6.8 & 15.9 & 396 & 180 & 80 
& 90 & 100 & 108.0 \\
\hline
web mail & 828 & 12.93 & 8.8 & 9 & 8.6 & 9.6 & 11.3 & 545 & 170 & 100 & 
90 & 90 & 108.7 \\
\hline
iphone & 200 & 8.9 & 1.3 & 1.2 & 1 & 2.7 & 37.3 & 600 & 160 & 90 & 80 & 
90 & 101.3 \\
\hline
jeux.fr & 1060 & 14.3 & 10.4 & 10.8 & 11 & 11.4 & 9.3 & 587 & 160 & 100 
& 90 & 90 & 106.7 \\
\hline
roland garros & 899 & 23.55 & 5 & 4.6 & 4.7 & 8.5 & 12.9 & 591 & 190 & 
90 & 90 & 90 & 110.0 \\
\hline
robert pattinson & 639 & 157.31 & 22 & 21 & 23 & 49.1 & 1.6 & 124 & 150 
& 60 & 70 & 60 & 80.7 \\
\hline
mappy michelin & 658 & 13.1 & 5.2 & 5.3 & 5.1 & 6.8 & 17.2 & 27 & 210 & 
90 & 110 & 80 & 116.7 \\
\hline
le monde & 1143 & 21.68 & 6.8 & 6.8 & 6.8 & 9.8 & 12.1 & 671 & 180 & 110 
& 90 & 110 & 118.7 \\
\hline
figaro & 617 & 15.7 & 4 & 3.9 & 3.8 & 6.3 & 20.3 & 674 & 250 & 120 & 90 
& 80 & 127.3 \\
\hline
tf1 & 385 & 9.54 & 2.5 & 2.6 & 2.5 & 3.9 & 33.0 & 680 & 290 & 110 & 80 & 
80 & 130.0 \\
\hline
le parisien & 605 & 7.04 & 4.3 & 4.4 & 4.3 & 4.9 & 20.7 & 643 & 170 & 90 
& 80 & 80 & 100.7 \\
\hline
liberation & 308 & 6.1 & 1.9 & 1.9 & 1.8 & 2.7 & 36.6 & 684 & 150 & 80 & 
80 & 100 & 99.3 \\
\hline
sarkozy & 800 & 7.08 & 5.7 & 6.1 & 5.9 & 6.1 & 16.8 & 642 & 170 & 90 & 
80 & 90 & 103.3 \\
\hline
20 minutes & 577 & 75.63 & 5.8 & 5.5 & 5.6 & 19.6 & 6.0 & 661 & 180 & 
130 & 100 & 80 & 118.7 \\
\hline
obama & 267 & 6.1 & 1.5 & 1.7 & 1.6 & 2.5 & 45.3 & 634 & 180 & 90 & 110 
& 90 & 113.3 \\
\hline
news & 800 & 9.71 & 6.7 & 6.5 & 6.5 & 7.2 & 15.3 & 657 & 150 & 100 & 100 
& 100 & 110.0 \\
\hline
les echos & 1509 & 82.03 & 8.5 & 9.3 & 8.3 & 23.4 & 4.6 & 645 & 180 & 90 
& 80 & 100 & 108.0 \\
\hline
pub orange & 1053 & 19.46 & 10 & 10.3 & 10.4 & 12.1 & 11.1 & 383 & 260 & 
100 & 90 & 120 & 134.7 \\
\hline
pub vittel & 581 & 33.65 & 4.2 & 4.3 & 4.6 & 10.2 & 11.6 & 36 & 180 & 
100 & 100 & 110 & 118.7 \\
\hline
pub tf1 & 782 & 7.2 & 7.3 & 7.2 & 7.2 & 7.2 & 17.3 & 453 & 210 & 90 & 
110 & 110 & 124.7 \\
\hline
pub sfr & 558 & 10.5 & 5.2 & 4.6 & 4.6 & 5.9 & 19.0 & 85 & 190 & 90 & 90 
& 100 & 112.7 \\
\hline
pub renault & 801 & 13.1 & 9.6 & 9.6 & 9.6 & 10.3 & 7.8 & 251 & 270 & 80 
& 10 & 10 & 80.7 \\
\hline
pub oasis & 800 & 8.08 & 7.9 & 8 & 7.9 & 8.0 & 16.0 & 114 & 250 & 90 & 
110 & 90 & 127.3 \\
\hline
pub nike & 635 & 6.9 & 5.7 & 5.6 & 5.6 & 5.9 & 21.9 & 105 & 190 & 90 & 
110 & 140 & 128.7 \\
\hline
pub iphone & 600 & 6 & 5 & 5.1 & 4.8 & 5.2 & 22.3 & 125 & 190 & 90 & 110 
& 90 & 115.3 \\
\hline
pub free & 755 & 24.6 & 4.1 & 4.1 & 3.9 & 8.1 & 14.8 & 504 & 230 & 100 & 
90 & 90 & 120.7 \\
\hline
pub evian & 486 & 13.9 & 3.8 & 3.9 & 3.9 & 5.9 & 21.6 & 67 & 220 & 90 & 
90 & 130 & 126.7 \\
\hline
twilight & 701 & 17.2 & 5.1 & 5.4 & 5.1 & 7.6 & 15.1 & 614 & 160 & 120 & 
90 & 100 & 114.7 \\
\hline
michael jackson & 1090 & 20.5 & 7.6 & 7.1 & 7 & 9.9 & 13.2 & 670 & 200 & 
110 & 100 & 130 & 130.7 \\
\hline
wat & 167 & 7.9 & 0.89 & 0.92 & 0.9 & 2.3 & 53.6 & 605 & 230 & 100 & 90 
& 100 & 123.3 \\
\hline
programme tnt & 692 & 20.94 & 8.2 & 8.2 & 12.72 & 12.0 & 10.2 & 491 & 
220 & 100 & 100 & 90 & 121.3 \\
\hline
naruto shippuden & 311 & 110.3 & 2.1 & 1.9 & 2.1 & 23.7 & 4.6 & 344 & 
180 & 90 & 90 & 90 & 108.0 \\
\hline
streaming & 250 & 5.6 & 1.4 & 1.3 & 1.3 & 2.2 & 46.6 & 422 & 150 & 90 & 
90 & 90 & 102.0 \\
\hline
m6 replay & 286 & 7.8 & 1.9 & 1.8 & 2 & 3.1 & 35.1 & 72 & 180 & 80 & 100 
& 90 & 108.0 \\
\hline
one piece & 658 & 11.2 & 4.7 & 4.8 & 4.6 & 6.0 & 21.6 & 618 & 220 & 110 
& 90 & 120 & 129.3 \\
\hline
Twitter & 191 & 2.9 & 1 & 1 & 0.9 & 1.4 & 87.7 & 618 & 220 & 100 & 90 & 
90 & 118.7 \\
\hline
Swine Flu & 190 & 10.94 & 1.1 & 1.2 & 1 & 3.1 & 32.4 & 69 & 190 & 80 & 
80 & 70 & 99.3 \\
\hline
Stock Market & 807 & 11.2 & 8.8 & 8.5 & 8.5 & 9.1 & 16.0 & 251 & 210 & 
120 & 130 & 140 & 146.0 \\
\hline
Farrah Fawcett & 251 & 27.5 & 1.5 & 1.5 & 1.4 & 6.7 & 16.4 & 83 & 200 & 
80 & 90 & 90 & 109.3 \\
\hline
Patrick Swayze & 1258 & 19.35 & 9.1 & 9.2 & 9.1 & 11.2 & 9.9 & 146 & 220 
& 80 & 70 & 100 & 110.7 \\
\hline
Cash for Clunkers & 400 & 18.43 & 2.8 & 2.8 & 2.8 & 5.9 & 14.2 & 2 & 140 
& 70 & 70 & 70 & 84.0 \\
\hline
Jon and Kate Gosselin & 1040 & 25.11 & 11.2 & 12 & 11.3 & 14.2 & 5.6 & 4 
& 120 & 70 & 70 & 70 & 80.0 \\
\hline
Billy Mays & 1153 & 142.7 & 7.5 & 7.2 & 7.3 & 34.4 & 3.2 & 83 & 190 & 90 
& 100 & 80 & 110.0 \\
\hline
Jaycee Dugard & 26 & 16.9 & 0.4 & 0.5 & 0.4 & 3.7 & 23.6 & 15 & 240 & 50 
& 50 & 50 & 88.0 \\
\hline
Jean Sarkozy & 1047 & 168.271 & 94 & 97 & 95 & 109.9 & 1.2 & 608 & 210 & 
100 & 120 & 100 & 127.3 \\
\hline
Rihanna & 123 & 2.8 & 0.7 & 0.7 & 0.6 & 1.1 & 120.1 & 587 & 270 & 110 & 
90 & 90 & 131.3 \\
\hline
Zohra Dhati & 53 & 2.8 & 0.5 & 0.4 & 0.4 & 0.9 & 119.1 & 5 & 140 & 90 & 
100 & 110 & 108.0 \\
\hline
Salma Hayek et François Pinault & 469 & 26.421 & 3.3 & 3.2 & 3.2 & 7.9 & 
9.9 & 8 & 190 & 50 & 50 & 50 & 78.0 \\
\hline
Frédéric Mitterrand & 1013 & 113.09 & 9 & 9.4 & 9.2 & 30.0 & 4.4 & 586 & 
200 & 130 & 110 & 100 & 130.7 \\
\hline
Roman Polanski & 468 & 16.02 & 2.9 & 2.9 & 2.8 & 5.5 & 26.9 & 591 & 260 
& 100 & 100 & 160 & 148.0 \\
\hline
Loana & 77 & 69.68 & 0.9 & 0.8 & 0.8 & 14.6 & 6.9 & 160 & 131 & 90 & 90 
& 100 & 100.9 \\
\hline
Caster Semenya & 106 & 74.58 & 1 & 0.8 & 0.9 & 15.6 & 5.2 & 44 & 170 & 
60 & 60 & 60 & 82.0 \\
\hline
Jacques Séguéla & 1144 & 27.86 & 20 & 20 & 19.9 & 21.5 & 7.1 & 140 & 310 
& 130 & 100 & 110 & 152.7 \\
\hline
Yann Barthes & 749 & 158.229 & 4.7 & 4.4 & 4.4 & 35.2 & 5.4 & 113 & 250 
& 290 & 120 & 120 & 191.3 \\
\hline
Le miracle de l'Hudson & 1400 & 125.08 & 17 & 19 & 17.6 & 39.3 & 2.8 & 
167 & 430 & 130 & 110 & 120 & 110.0 \\
\hline
Barack Obama & 459 & 14.23 & 2.4 & 2.6 & 2.4 & 4.8 & 24.8 & 615 & 170 & 
90 & 120 & 110 & 119.3 \\
\hline
La crise économique & 1076 & 15.6 & 6.3 & 6.9 & 6.3 & 8.3 & 16.8 & 700 & 
220 & 130 & 120 & 110 & 140.0 \\
\hline
Greve aux Antilles & 952 & 30.54 & 7.5 & 7.5 & 8.6 & 12.4 & 13.9 & 248 & 
300 & 150 & 120 & 150 & 172.0 \\
\hline
Séisme en Italie & 1000 & 18.75 & 8.6 & 9.1 & 8.6 & 10.8 & 11.5 & 549 & 
180 & 90 & 90 & 150 & 124.0 \\
\hline
La grippe A & 631 & 46.93 & 5 & 4 & 4 & 12.9 & 8.9 & 612 & 200 & 100 & 
90 & 90 & 114.7 \\
\hline
Le malaise présidentiel & 799 & 17.9 & 8.3 & 8.1 & 8.5 & 10.2 & 13.4 & 
204 & 230 & 110 & 120 & 110 & 136.7 \\
\hline
Hadopi & 225 & 27.05 & 1.1 & 1 & 1.1 & 6.3 & 18.9 & 438 & 220 & 90 & 90 
& 100 & 118.7 \\
\hline
Le proces Clearstream & 517 & 31.5 & 5.6 & 5.9 & 5.6 & 10.9 & 12.0 & 109 
& 240 & 110 & 100 & 100 & 130.7 \\
\hline
Madonna & 800 & 20.25 & 6.5 & 6.3 & 6.3 & 9.1 & 13.1 & 641 & 200 & 100 & 
100 & 100 & 120.0 \\
\hline
U2 & & 11.35 & 2.9 & 3.1 & 2.8 & 4.6 & 22.7 & 602 & 150 & 90 & 90 & 100 
& 104.7 \\
\hline
Diam's & 142 & 16.58 & 0.73 & 0.78 & 0.74 & 3.9 & 28.9 & 378 & 220 & 80 
& 90 & 90 & 113.3 \\
\hline
Mylene Farmer & 540 & 24.45 & 3 & 3.1 & 2.9 & 7.3 & 14.3 & 586 & 160 & 
90 & 90 & 90 & 104.0 \\
\hline
Les Beatles remastérisés & 520 & 19.56 & 3.7 & 4.1 & 3.7 & 7.0 & 23.7 & 
110 & 320 & 140 & 120 & 120 & 165.3 \\
\hline
Johnny Hallyday & 822 & 8.8 & 5.6 & 6 & 5.5 & 6.3 & 23.7 & 535 & 270 & 
110 & 130 & 120 & 150.0 \\
\hline
Lady Gaga & 489 & 74.68 & 3.8 & 3.7 & 3.7 & 17.9 & 6.6 & 579 & 220 & 90 
& 90 & 100 & 118.7 \\
\hline
La séparation d'Oasis & 1360 & 28.62 & 15.7 & 15.7 & 15.7 & 18.3 & 10.1 
& 207 & 280 & 260 & 100 & 120 & 184.0 \\
\hline
Prince a Paris & 800 & 9.9 & 7.7 & 7.7 & 7.7 & 8.1 & 20.8 & 679 & 260 & 
180 & 120 & 140 & 169.3 \\
\hline
David Guetta & 736 & 109.37 & 27 & 26 & 24 & 42.4 & 2.8 & 529 & 200 & 90 
& 100 & 100 & 117.3 \\
\hline
 & & & & & & & & & & & & & \\
\hline
 & & 30.5 & 6.8 & 6.8 & 6.8 & 11.6 & 20.5 & & 207.0 & 101.8 & 94.3 & 98.2 
& 119.1 \\
\hline
\end{tabular}
\end{table}

\begin{table}[h]
\centering
\begin{tabular}{|p{4cm}|*{13}{p{0.5cm}|}}
\hline
Wikipedia En & \multicolumn{6}{p{3cm}|}{mARC Search Engine demonstrator} & \multicolumn{1}{p{0.5cm}|}{\textbf{Ratio}} & \multicolumn{6}{p{3cm}|}{Google} \\
\hline
 &Results (real)&Req1 (ms) out of cache&Req2 (ms)&Req3 
(ms) &Req3 (ms)&average (ms)&&Results(real)& 
Req1(ms) out of cache&Req2 (ms)&Req3 (ms)&Req 3 (ms)& 
average (ms) \\
\hline
Mathematical formulations of quantum mechanics & 1360 & 55.1 & 8.6 & 8.7 
& 8.7 & 18.0 & 10.2 & 581 & 280 & 160 & 160 & 160 & 184.0 \tabularnewline
\hline
Philosophical interpretation of classical physics & 1284 & 51.1 & 8.9 & 
8.6 & 8.9 & 17.3 & 11.5 & 536 & 350 & 160 & 160 & 160 & 198.0 \\
\hline
Governor General's Award for English language non fiction & 893 & 77.4 & 
5.4 & 5.7 & 5.3 & 19.9 & 5.8 & 545 & 220 & 90 & 90 & 90 & 116.0 \\
\hline
John Breckinridge (Attorney General) & 1145 & 42.97 & 7.5 & 7.5 & 7.5 & 
14.6 & 12.2 & 540 & 300 & 130 & 150 & 160 & 177.3 \\
\hline
Popular Front for the Liberation of Palestine General Command & 959 & 
73.2 & 5.9 & 6.4 & 5.9 & 19.5 & 6.7 & 484 & 230 & 120 & 110 & 90 & 131.3 
\\
\hline
The Six Wives of Henry VIII (TV series) & 891 & 18.93 & 6.4 & 6 & 5.9 & 
8.7 & 20.2 & 449 & 370 & 140 & 110 & 130 & 175.3 \\
\hline
List of Chancellors of the University of Cambridge & 1060 & 17.35 & 9.2 
& 8.6 & 8.6 & 10.5 & 15.9 & 592 & 300 & 140 & 150 & 110 & 166.7 \\
\hline
International Council of Unitarians and Universalists & 1045 & 48.44 & 8 
& 7.8 & 7.8 & 16.0 & 8.7 & 573 & 270 & 110 & 110 & 100 & 139.3 \\
\hline
International Council of Unitarians and Universalists & 739 & 232.4 & 
4.6 & 4.7 & 4.3 & 50.1 & 2.6 & 362 & 280 & 90 & 90 & 90 & 128.0 \\
\hline
Finitely generated abelian group & 655 & 29.3 & 3.3 & 3.4 & 3.3 & 8.5 & 
21.6 & 288 & 240 & 170 & 170 & 170 & 184.0 \\
\hline
Structure theorem for finitely generated modules over a principal ideal 
domain & 495 & 102.11 & 2.4 & 2.4 & 2.4 & 22.3 & 10.7 & 69 & 310 & 210 & 
220 & 230 & 238.0 \\
\hline
Asimov's Biographical Encyclopedia of Science and Technology & 1401 & 
93.15 & 8.4 & 8.8 & 8.4 & 25.5 & 7.2 & 271 & 260 & 160 & 160 & 170 & 
182.7 \\
\hline
Aalto University School of Science and Technology & 974 & 19.59 & 6 & 
5.9 & 5.9 & 8.7 & 13.2 & 180 & 250 & 80 & 80 & 80 & 114.0 \\
\hline
List of historical sites associated with Ludwig van Beethoven & 604 & 
63.97 & 3.4 & 3 & 3.1 & 15.3 & 11.0 & 104 & 270 & 140 & 150 & 140 & 
168.7 \\
\hline
In France , the President of the General Council ( French language 
French : ''Président du conseil général'') is the locally-elected head 
of the General councils of France General Council , the assembly 
governing a Departments of France department & 707 & 104.6 & 7.39 & 5 & 
5.6 & 25.7 & 35.1 & 16 & 910 & 870 & 920 & 910 & 902.0 \\
\hline
The cinema of the Soviet Union , not to be confused with Cinema of 
Russia despite Russian language films being predominant in both genres, 
includes several film contributions of the constituent republics of the 
Soviet Union reflecting elements of & 600 & 141.53 & 5.4 & 5.7 & 4.7 & 
32.5 & 8.4 & 23 & 760 & 140 & 170 & 150 & 274.7 \\
\hline
Niger is home to a number of national parks and protected areas , 
including two UNESCO-MAB Biosphere Reserves. The protected areas of 
Niger normally have a designation and status determined by the 
Government of Niger. & 715 & 75.6 & 5.4 & 4.7 & 4.7 & 19.1 & 32.9 & 4 & 
860 & & 860 & 850 & 628.0 \\
\hline
The term hamburger or burger can also be applied to the patty meat patty 
on its own, especially in the UK. There are several accounts of the 
invention of the hamburger & 672 & 71.32 & 4.6 & 4.1 & 4.5 & 17.8 & 12.3 
& 132 & 590 & 130 & 120 & 130 & 219.3 \\
\hline
Rockwell International was a major American manufacturing conglomerate 
(company) conglomerate in the latter half of the 20th century, involved 
in aircraft, the space industry, both defense-oriented and commercial 
electronics, automotive and truck & 656 & 63.81 & 5 & 4.7 & 4.9 & 16.7 & 
38.2 & 1 & 850 & 810 & 130 & 810 & 636.7 \\
\hline
 & & & & & & & & & & & & & \\
\hline
 & & 72.7 & 6.1 & 5.9 & 5.8 & 19.3 & 15.0 & & 415.8 & 213.9 & 216.3 & 
248.9 & 261.3 \\
\hline
\end{tabular}
\end{table}

\begin{table}[h]
\centering
\begin{tabular}{|p{3cm}|*{13}{p{0.5cm}|}}
\hline
Wikipedia Fr & \multicolumn{6}{p{3cm}|}{mARC Search Engine demonstrator} & \multicolumn{1}{p{0.5cm}|}{\textbf{Ratio}} & \multicolumn{6}{p{3cm}|}{Google} \\
\hline
 & & & \\
\hline
 &Results (real)&Req1 (ms) out of cache&Req2 (ms)&Req3 
 (ms)&Req3 (ms)& average (ms)&&Results (real)& 
Req1 (ms) out of cache&Req2 (ms)&Req3 (ms)&Req3 (ms)& 
average (ms) \\
\hline
Les Monstres du fond des mers & 794 & 10.853 & 4.7 & 4.7 & & 5.7 & 22.7 
& 482 & 220 & 100 & 110 & 110 & 129.3 \\
\hline
Liste des lacs et mers intérieures de la Terre du Milieu & 504 & 31.17 & 
3.8 & 3.9 & 3.7 & 9.3 & 22.6 & 419 & 420 & 130 & 170 & 170 & 209.3 \\
\hline
Liste des lacs de Suisse par canton & 581 & 33.22 & 4.1 & 3.9 & 4.1 & 
9.9 & 13.9 & 342 & 220 & 120 & 110 & 120 & 137.3 \\
\hline
Liste d'archéologues par ordre alphabétique & 1474 & 30.53 & 9.4 & 8.8 & 
9.4 & 13.5 & 11.1 & 556 & 270 & 150 & 100 & 110 & 150.0 \\
\hline
Liste des noms de famille les plus courants au Québec, par ordre 
alphabétique H & 1399 & 39.9 & 10.6 & 12.3 & 10.5 & 16.9 & 7.2 & 584 & 
250 & 100 & 90 & 80 & 122.0 \\
\hline
Moulin a vent de l'île Saint Bernard de Châteauguay & 802 & 49.37 & 5.7 
& 5.7 & 5.7 & 14.4 & 16.0 & 200 & 330 & 210 & 200 & 210 & 231.3 \\
\hline
César de la meilleure actrice dans un second rôle & 1227 & 68.334 & 11.7 
& 11.2 & 11.8 & 22.9 & 7.4 & 546 & 280 & 110 & 210 & 110 & 170.7 \\
\hline
La littérature française comprend l'ensemble des oeuvres écrites par des 
auteurs de nationalité française ou de langue française . Son histoire 
commence en ancien français au Moyen Age et se perpétue aujourd'hui. 
Chanson de geste La Littér & 556 & 86.47 & 6.9 & 6.1 & 6.1 & 22.4 & 20.6 
& 7 & 970 & 540 & 300 & 160 & 460.7 \\
\hline
Un roton est une quasiparticule , un quantum d'excitation de l' hélium 
superfluide , avec des propriétés et notamment un spectre différent de 
celui des phonon & 380 & 43.1 & 2.8 & 2.7 & 2.7 & 10.8 & 72.9 & 1 & 820 
& 740 & 840 & 760 & 788.0 \\
\hline
La résonance est un phénomene selon lequel certains systemes physiques 
(électriques, mécaniques...) sont sensibles a certaines fréquences. Un 
systeme résonant peut accumuler une énergie, si celle-ci est appliquée 
sous forme périodique, et proche d'une fr & 653 & 20.66 & 6.2 & 6.2 & 
6.2 & 9.1 & 67.0 & 3 & 1060 & 930 & 290 & 270 & 609.3 \\
\hline
Brevet de technicien supérieur Techniques physiques pour l'industrie et 
le laboratoire & 450 & 47.6 & 3.7 & 3.8 & 3.8 & 12.5 & 13.5 & 115 & 310 
& 140 & 130 & 130 & 168.7 \\
\hline
Diplôme universitaire de technologie Mesures physiques & 519 & 47.32 & 
3.7 & 3.6 & 3.7 & 12.4 & 14.0 & 340 & 310 & 160 & 140 & 120 & 174.0 \\
\hline
 & & & & & & & & & & & & & \\
\hline
 & & 42.4 & 6.1 & 6.1 & 6.2 & 13.3 & 24.1 & & 455.0 & 285.8 & 224.2 & 
195.8 & 279.2 \\
\hline
\end{tabular}
\end{table}

\end{document}